\documentclass[12pt]{article}
\usepackage{graphicx}
\usepackage{amsmath}
\newcommand{\sm}{\hbox{$\bigcirc$\kern-0.72em\hbox{\bf s} }}
\newcommand{\Id}{\hbox{\sl 1\kern-0.25em\hbox{I} }}
\newcommand{\rcorr}{\hbox{\kern-1.2em$\longrightarrow$}}
\newcommand{\lrcorr}{\hbox{\kern-1.2em$\longleftrightarrow$}}

\newcommand{\nRightarrow}{\Rightarrow\kern-1.2em\hbox{/}\kern.8em} %
\newcommand{\BB}{\hbox{I\kern-.2em\hbox{B}}} %BB
\newcommand{\DD}{\hbox{I\kern-.2em\hbox{D}}} %DD
\newcommand{\FF}{\hbox{I\kern-.2em\hbox{F}}} %FF
\newcommand{\NN}{\hbox{I\kern-.2em\hbox{N}}}  %Naturali
\newcommand{\ZZ}{{{\rm Z}\kern-.28em{\rm Z}}} %Interi
\newcommand{\RR}{\mathop{{\rm I}\kern-.2em{\rm R}}\nolimits} %Reali
\newcommand{\RRe}{\mathop{{\rm I}\kern-.2em{\rm Re}}\nolimits} %Reali
\newcommand{\QQ}{\hbox{l\kern-.36em\hbox{Q}}}  %Razionali
\newcommand{\CC}{\hbox{{\textsf I}\kern-.47em\hbox{C}}}
\newcommand{\nop}{\hbox{{\textsf I}\kern-.47em\hbox{O}}}

\def\TREV{{{}^\triangleleft\kern-1.5pt\texttt{T}}}
\def\trev{{{}^\triangleleft\kern-3.2pt\texttt{t}}}
\def\SREV{{{}_\triangleleft\kern-2pt\texttt{S}}}
\def\srev{{{}_\triangleleft\kern-2.2pt\texttt{s}}}
\begin{document}
\title{Group theoretical derivation of\\ consistent massless particle theories}
\author{Giuseppe Nistic\`o
\\
{\small Dipartimento
di Matematica e Informatica, Universit\`a della Calabria, Italy}\\
{\small and}\\
{\small
INFN -- gruppo collegato di Cosenza, Italy}\\
{\small email: giuseppe.nistico@unical.it} } \maketitle
%%%%%%%%%%%%%%%%%%%%%%%%%%%%%% ABSTRACT
\abstract{Current theories of massless free particle assume {\sl  unitary} space inversion and {\sl anti-unitary} 
time reversal operators. In so doing robust classes of possible theories are discarded.
In the present work theories of massless systems are derived through a strictly deductive
development from the principle of relativistic invariance, so that a kind of space inversion or 
time reversal operator is ruled out  only if it causes inconsistencies.
As results, new classes of consistent theories for massless isolated systems are explicitly determined.
On the other hand, the approach determines definite constraints implied by the invariance principle; 
they were ignored by some past investigations that, as a consequence, turn out to be not consistent 
with the invariance principle.
Also the problem of the {\sl  localizability} for massless systems is reconsidered within the
new theoretical framework, obtaining a generalization 
and a deeper detailing of previous results. 
}
%
%%%__________________________________________________________________________________________________  1
\section{Introduction}
Relativistic quantum theories of single free particle have been derived  by exploiting the implications of {\sl relativistic invariance},
the principle that establishes the invariance of the physical theory of an isolated system
with respect to relativistic changes of inertial reference frames
\cite{Wigner1},\cite{WignerBargmann},\cite{Wightman},\cite{Costa}.
The basic implication states that each symmetry transformation can be assigned a unitary or anti-unitary operator
able to realize the correponding transformation in the quantum theory of 
the system according to Wigner relation \cite{WignerBook}. Moreover,
such an assignment gives rise to  a unitary representation $U$ of
$\tilde{\mathcal P}_+^\uparrow$, the universal covering group of the proper orthochronus Poincar\'e group 
${\mathcal P}_+^\uparrow$.
Also space inversion and time reversal  are transformations that should be not excluded {\it apriori} 
from the class of symmetry transformations. So, the structure of the
quantum theory of an isolated system must contain a tern $(U,\SREV,\TREV)$ formed by a unitary 
representation $U$ of $\tilde{\mathcal P}_+^\uparrow$, and by two operators $\SREV$ and $\TREV$, each of them unitary or anti-unitary,
which can realize {\sl all} quantum transformations associated to elements of Poincar\'e group.
\par
Contributions from the literature \cite {Wigner1},\cite{Weinberg1},\cite{WeinbergBook},\cite{Jordan1}
in general discard terns with $\SREV$ anti-unitary or $\TREV$ unitary \cite{Wightman},
so that  the class of the possible quantum theories of isolated system undergoes a drastic reduction.
One reason put forward for this exclusion \cite{WeinbergBook},\cite{Costa},\cite{NiedererAl} 
is that  anti-unitarity space inversion operators, or unitary time reversal operators, change the sign of the spectral 
values of the self-adjoint generator $P_0$ of time translations, that in these studies is identified with {\sl energy}, 
and it is usually assumed that this change cannot happen (see \cite{NiedererAl}, p. 135).
However, in \cite{Nistico3} it is shown that for particles with ``non-zero'' mass, perfectly consistent theories 
can be derived from the invariance principle,  characterized by anti-unitary space inversion 
operator $\SREV$ or unitary time reversal operator $\TREV$;
in particular, theories for Klein-Gordon type particle are identified, without the inconsistencies suffered by the
early theories \cite{Klein},\cite{Fock},\cite{Gordon}.
\par 
In the present work consistent  quantum theories of  ``massless'' isolated systems
are derived from the invariance principle, through a strictly deductive development to avoid
{\it apriori} preclusions about the unitary or anti-unitary character of $\SREV$
or $\TREV$. Obviously, this methodological commitment is incompatible with a quantum field theory approach.
A kind of space inversion or time reversal operator will be ruled out  only if it
causes inconsistencies; on the other hand, definite constraints for the tern $(U,\SREV,\TREV)$ are determined,
ignored by past investigations \cite{Weinberg1},\cite{Jordan1}.
As results, new classes of consistent theories for massless isolated systems are explicitly determined,
while some previous theory turns out to be inconsistent with the invariance principle.
\par
The present extension
requires to reconsider the {\sl  localizability} problem for massless systems,
that is to say the problem of ascertaining whether a unique {\sl position operator} exists or not within each
possible theory for massless systems.
The investigations about this problem
trace back to the paper of Newton and Wigner \cite{NewtonWigner}, but the problem was addressed by many researchers following 
different approaches, e.g. in \cite{Wightman},{\cite{Amrein},{\cite{Jordan1}.
We address the localizability problem in the present extended theoretical framework, obtaining a generalization 
and a deeper detailing of previous results. 
\vskip1pc
The work makes use of basic 
 mathematical notions and results outlined in section 2.
In section 3.1 and 3.2
the existence of a transformer tern $(U,\SREV,\TREV)$ is derived within
general implications that hold for every isolated system.
In section 3.3  the further conditions are identified to be obeyed by the theory of an isolated system in order
to be the theory of a massless elementary free particle, such as the irreducibility of the tranformer tern of the theory.
\par
Since unbounded helicities have never been observed in Nature,
the irreducibility condition indicates that the identification of the empirically meaningful possible theories of a massless elementary
particle entails the identification of the class ${\mathcal I}$ of all irreducible transformer terns with bounded helicity.
Section 4 
classifies these irreducible terns into three subclasses 
${\mathcal I}({\bf u})$, ${\mathcal I}({\bf d})$, ${\mathcal I}({\bf s})$; each of them turns out to be characterized
by specific combination of the unitary or anti-unitary characters of $\SREV$ and $\TREV$.
\par
In section 5 the terns of the classes ${\mathcal I}({\bf u})$ and ${\mathcal I}({\bf d})$ 
are identified and investigated.
It is found in particular that all terns are constrained to have zero helicity.
\par
In section 6 the empirically meaningful irreducible terns of the class ${\mathcal I}({\bf s})$ are identified. Consistent theories are
found with anti-unitary $\SREV$ or unitary $\TREV$, ignored by the past literature.
Among the results, it is found that
irreducible terns with non-zero but opposite  values of the helicity exist in this class.
\par
Hence, the class of the possible theories of massless isolated system turns out to be 
structurally modified: it is extended to admit theories with  anti-unitary $\SREV$ and unitary $\TREV$.
This modification requires to re-consider the problem of the localizability of massless particles. This task is
addressed in section 7. 
The localizability of massless particle with zero helicity is proved to hold also for particles with terns 
in  ${\mathcal I}({\bf s})$; three inequivalent theories are explicitly identified.
With regard to systems with non-zero helicity, past investigations proved non-localizabity
by making use of terns in ${\mathcal I}({\bf u})$ or ${\mathcal I}({\bf d})$; but section 5 proved that
helicity is zero in these cases, so that the proofs are not effective \cite{Jordan1}.
In section 7.2 an alternative general proof of non-localizability is given without these shortcomings. 

%
%%  ___________________________________________________________________________________________                 2                                                                                                                                                                                                          
\section{Starting notions}
In this section theoretical and mathematical concepts are reviewed, necessary to the development of  the work.
%%%%%___________________________________________________________________________________________  2.1
\subsection{General quantum formalism}
The quantum theory of a
physical system, formulated in a complex and separable Hilbert space $\mathcal H$, is based on 
the following mathematical structures.
\begin{description}
\item[\quad-]
The set $\Omega({\mathcal H})$ of all self-adjoint operators of $\mathcal H$, which represent
{\sl quantum observables}.
\item[\quad-]
The complete, ortho-complemented lattice $\Pi({\mathcal H})$ of all projections operators of $\mathcal H$, i.e.
quantum observables with possible outcomes
in $\{0,1\}$.
\item[\quad-]
The set $\Pi_1({\mathcal H})$ of all rank one orthogonal projections of $\mathcal H$.
\item[\quad-]
The set ${\mathcal S}({\mathcal H})$ of all density operators of $\mathcal H$,
which represent {\sl quantum states}.
\item[\quad-]
The set ${\mathcal V}({\mathcal H})$ of all unitary  or anti-unitary operators of the Hilbert space $\mathcal H$.
\item[\quad-]
The set ${\mathcal U}({\mathcal H})$ of all unitary operators of $\mathcal H$; trivially,
${\mathcal U}({\mathcal H})\subseteq{\mathcal V}({\mathcal H})$ holds.
\end{description}
%%____________________________________________________________________________________________________  2.2
\subsection{Poincar\'e groups and representations}
Given any vector $\underline x=(x_0,{\bf x})\in\RR^4$, we call $x_0$ the {\sl time component} of $\underline x$
and ${\bf x}=(x_1,x_2,x_3)$ the {\sl spatial component} of $\underline x$.
The proper orthochronous Poincar\'e group ${\mathcal P}_+^\uparrow$ is the separable locally compact
group of all transformations of $\RR^4$
generated by the ten one-parameter sub-groups ${\mathcal T}_0$, ${\mathcal T}_j,{\mathcal R}_j$, ${\mathcal B}_j$, $j=1,2,3$,
of time translations, spatial translations, proper spatial rotations and Lorentz boosts, respectively, relative to the axes $x_j$.
The Euclidean group $\mathcal E$ is the sub-group generated by all ${\mathcal T}_j$ and ${\mathcal R}_j$.
The sub-group generated by all
${\mathcal R}_j$, ${\mathcal B}_j$ is the proper orthochronous Lorentz group ${\mathcal L}_+^\uparrow$ \cite{BarutBook}.
It does not include time reversal $\trev$ and space inversion $\srev$.
Time reversal $\trev$ transforms $\underline x=(x_0,{\bf x})$ into $(-x_0,{\bf x})$; space inversion $\srev$
transforms $\underline x=(x_0,{\bf x})$ into $(x_0,-{\bf x})$.
The group generated by $\{{\mathcal P}_+^\uparrow, \trev,\srev\}$ is the separable and locally compact
Poincar\'e group $\mathcal P$. 
\par
By ${\mathcal L}_+$ we denote the subgroup generated by ${\mathcal L}_+^\uparrow$ and $\trev$, while ${\mathcal L}^\uparrow$
denotes the subgroup generated by ${\mathcal L}_+^\uparrow$ and $\srev$; analogously, ${\mathcal P}_+$ denotes
the subgroup generated by ${\mathcal P}_+^\uparrow$ and $\trev$, while ${\mathcal P}^\uparrow$ is the subgroup
generated by ${\mathcal P}_+^\uparrow$ and $\srev$.
\par
The subgroup of $\mathcal P$  generated by $\srev$ and $\trev$ is ${\mathcal D}_0=\{e,\srev,\trev,\trev\srev\}$ .
Since ${\mathcal D}_0\cap{\mathcal P}_+^\uparrow=\{e\}$, every $g\in\mathcal P$ is the product
$g=g_0g_1$ of a {\sl unique} pair $(g_0,g_1)\in{\mathcal D}_0\times{\mathcal P}_+^\uparrow$.
\par
In our investigation an important role is played by the semidirect product $\tilde{\mathcal P}_+^\uparrow=\RR^4\sm  SL(2,\CC)$
of the additive group $\RR^4$ and the group $SL(2,\CC)=\{\underline\Lambda\in GL(2,\CC)\mid \det\underline\Lambda=1\}$.
$\tilde{\mathcal P}_+^\uparrow$ is the universal covering group of ${\mathcal P}_+^\uparrow$. 
Accordingly, $\tilde{\mathcal P}_+^\uparrow$ is simply connected and 
there is a canonical homomorphism
${\textsf h}:\tilde{\mathcal P}_+^\uparrow\to\mathcal P$, $g\to {\textsf h}(g)\in\mathcal P$
that becomes an isomorphism when restricted to a suitable neighborhood of the identity
$(0,\Id_{_{\CC^2}})$ of $\tilde{\mathcal P}_+^\uparrow$.
\par
We denote by  $\tilde{\mathcal T}_0$, $\tilde{\mathcal T}_j,\tilde{\mathcal R}_j$, $\tilde{\mathcal B}_j$, 
$\tilde{\mathcal L}_+^\uparrow$ the subgroups of $\tilde{\mathcal P}_+^\uparrow$ that correspond to the subgroups
${\mathcal T}_0$, ${\mathcal T}_j,{\mathcal R}_j$, ${\mathcal B}_j$, 
${\mathcal L}_+^\uparrow$  of
${\mathcal P}_+^\uparrow$, respectively, through the homomorphism $\textsf h$.
All  one-parameter abelian subgroups $\tilde{\mathcal T}_0$, $\tilde{\mathcal T}_j,\tilde{\mathcal R}_j$, $\tilde{\mathcal B}_j$  of 
$\tilde{\mathcal P}_+^\uparrow$ are additive;
in fact, $\tilde{\mathcal B}_j$ is not additive with respect to the parameter
{\sl relative velocity} $u$, but it is additive with respect to the parameter $\varphi(u)=\frac{1}{2}\ln\frac{1+u}{1-u}$.
\vskip.5pc
The following definition collects general notions concerning group representations.
\vskip.5pc\noindent
{\bf Definition 2.1.} {\sl Let $G$ be a separable, locally compact group with identity element $e$. A correspondence
$U:G\to{\mathcal V}({\mathcal H})$, $g\to U_g$, with $U_e=\Id$, is a generalized projective representation of $G$
if the following conditions are satisfied.
\begin{description}
\item[\;{\rm i)}\;\;]
A complex function $\sigma:{G}\times{G}\to{\CC}$,
called multiplier, exists such that
\item[\;\;\quad] $U_{g_1g_2}=\sigma(g_1,g_2)U_{g_1}U_{g_2}$; the modulus $\vert\sigma(g_1,g_2)\vert$ is always 1, of course;
\item[\;{\rm ii)}\;]
for all $\phi,\psi\in\mathcal H$, the mapping $g\to\langle U_g\phi\mid\psi\rangle$ is a Borel function in $g$.
\end{description}\noindent
If $U_g$ is unitary for all $g\in G$, then $U$ is called
projective representation, or $\sigma$-representation; if also $\sigma(g_1,g_2)=1$ holds for all $g_1,g_2$, $U $ reduces to a (ordinary) unitary 
epresentation.
\par
A generalized projective representation is said to be continuous if for any fixed $\psi\in\mathcal H$
the mapping $g\to U_g\psi$ from $G$ to $\mathcal H$ is continuous with respect to $g$.
\par
If $U:G\to{\mathcal V}({\mathcal H})$ is a generalized projective representation and $\theta(g)\in\RR$ for all $g\in G$, with $e^{i\theta(e)}=\Id$, 
then the generalized projective representation $\tilde U:G\to{\mathcal V}({\mathcal H})$,
$g\to \tilde U_g=e^{i\theta(g)}U_g$ is said {\sl equivalent} to $U:G\to{\mathcal V}({\mathcal H})$.}
\vskip.5pc\noindent
In \cite{Nistico1} we have proved that the following statement holds.
\vskip.5pc\noindent
{\bf Proposition 2.1.}
{\sl Let $G$ be a separable locally compact group and let $U:G\to{\mathcal V}({\mathcal H})$ be a 
continuous generalized projective representation of $G$.
\begin{description}
\item[\;{\rm i)}\;\;]
If $G$ is a connected group, then $U$ is a projective
representation, i.e. $U_g\in{\mathcal U}({\mathcal H})$, for all $g\in G$.
\item[\;{\rm ii)}\;]
If $G$ is connected and simply connected, then then the phase exponent $\theta(g)$ can be chosen so that $U$ is equivalent to
a continuous ordinary unitary representation of $G$.
\end{description}
}
%
%%____________________________________________________________________________________________________   3
\section{General implications of Poincar\'e invariance}
Let  $\mathcal F$ be the class  of the (inertial) reference frames that move uniformly with respect to each other.
The following statement establishes the physical principle that characterizes an isolated system.
\begin{description}
\item[$\mathcal Sym$]{\sl
The theory of an isolated system is invariant with respect to changes of frames within the class $\mathcal F$.}
\end{description}
In section 3.1 it is outlined how this invariance principle implies the existence of a transformation $S_g$ acting on quantum obervables,
called {\sl quantum transformation}, in correspondence with each $g\in\mathcal P$. The structure of the quantum theory of 
an isolated system is then identified in a {\sl transformer tern}, whose mathematical properties of interest are stated in section 3.2.
In section 3.3 the concept of massless elementary free particle is introduced as a particular isolated localizable system.

%%_____________________________________________________________________________________________      3.1
\subsection{Quantum transformations}
Given a frame $\Sigma$ in $\mathcal F$ and $g\in\mathcal P$, by $\Sigma_g$ we denote the frame related to $\Sigma$ by such $g$.
Let ${\mathcal M}_A$ denote a procedure to perform the measurement of an observable represented by the self-adjoint 
operator $A$. The principle $\mathcal Sym$ implies that another
measuring procedure ${\mathcal M}'$ must exist, which is with respect to $\Sigma_g$ identical to what is ${\mathcal M}_A$
with respect to $\Sigma$, otherwise the invariance established by $\mathcal Sym$ would fail.
Then we denote by ${\it S_g}[A]$ the self-adjoint operator that represents the observable measured by ${\mathcal M}'$.
In so doing for every $g\in\mathcal P$ the mapping
$$
S_g:\Omega({\mathcal H})\to\Omega({\mathcal H})\,,\quad A\to S_g[A]\,
$$ 
is defined,
called {\sl quantum transformation associated to $g$}.
In \cite{Nistico1} it is proved that the following properties (S.1)-(S.3) of every quantum transformation 
 can be implied from the principle $\mathcal Sym$.
\begin{itemize}
\item[(S.1)] {\sl Every $S_g:\Omega({\mathcal H})\to\Omega({\mathcal H})$ is bijective;}
\item[(S.2)] {\sl for every $A\in\Omega({\mathcal H})$ and every real function $f$ such that  $f(A)\in\Omega({\mathcal H})$,
\item[] the equality $S_g[f(A)]=f(S_g[A])$ holds.}
\item[(S.3)] {\sl $S_{gh}[A]=S_g\left[S_h[A]\right]$, for all $g,h\in\mathcal P$ and every $A\in\Omega({\mathcal H})$.}
\end{itemize}
To every element $\tilde g$ of the covering group $\tilde{\mathcal P}_+^\uparrow$ we can associate the quantum transformation
$S_{\textsf h(\tilde g)}$  through the canonical homomorphism $\textsf h$.
Without introducing ambiguity, $S_{\textsf h(\tilde g)}$  can be denoted by $S_{\tilde g}$.
\par
Following \cite{Nistico1}, for any $S_ {\tilde g}$ a unitary or anti-unitary operator ${\tilde U}_{\tilde g}$ must exist
such that  $S_{\tilde g}[A]= \tilde U_{\tilde g}A\tilde U_{\tilde g}^{-1} =e^{i\theta(\tilde g)}\tilde U_{\tilde g}A\tilde(e^{i\theta(\tilde g)} U_{\tilde g})^{-1}$,
where $\theta$ is an arbitrary real function. Since ${\textsf h}$ is  a surjective homomorphism,
$S_{\tilde g_1\tilde g_2}[A]=S_{\tilde g_1}\left[S_{\tilde g_2}[A]\right]$ holds, and this implies that 
$\tilde g\to e^{i\theta(\tilde g)}\tilde U_{\tilde g}$ is a generalized projective representation, 
provided tha $ e^{i\theta(e)}\tilde U_{e}=\Id$,
$e$ being the neutral element of $\tilde{\mathcal P}_+^\uparrow$.
\par
The idea that small changes of reference frame, i.e. small $\tilde g$, cause small transformations
$A\to S_{\tilde g}[A]$ motivates the assumption that the mapping $\tilde g\to S_{\tilde g}$ is continuous according to Bargmann's
topology \cite{Nistico1}. Then, according to \cite{Nistico1},
the function $\theta$ can be chosen so that $\tilde g\to  e^{i\theta(\tilde g)}\tilde U_{\tilde g}$
is continuous. Therefore, by Prop.2.1 a further change of $\theta$ makes $\tilde g\to  e^{i\theta(\tilde g)}\tilde U_{\tilde g}\equiv U_{\tilde g}$
a continuous unitary projective representation of $\tilde{\mathcal P}_+^\uparrow$.
\par
Since the space inversion $\srev$ and the time reversal $\trev$  are  transformations 
not connected with the identity transformation $e\in\mathcal P$, 
the operator $\SREV$ and $\TREV$ that realize $S_\srev$ and $S_\trev$
can be unitary or anti-unitary.
\vskip.5pc
Thus,
the principle $\mathcal Sym$ ultimately implies that the quantum theory of an isolated system must admit a tern $(U,\SREV,\TREV)$, called 
the {\sl transformer tern of the theory}, formed by a continuous representation 
$U$ of $\tilde{\mathcal P}_+^\uparrow$ and by two operators $\SREV$, $\TREV$ such that
$$
S_{\tilde g}[A]=U_{\tilde g}AU_{\tilde g}^{-1}\,,\quad
\SREV A\SREV^{-1}=S_\srev[A]\,,\quad\TREV A\TREV^{-1}=S_\trev[A]
\eqno(1)
$$

%_________________________________________________________________________________________________    3.2
\subsection{The algebra of generators; helicity}
According to Stone's theorem \cite{Stone}, for any continuous unitary representation of ${\mathcal P}_+^\uparrow$
ten {\sl self-adjoint} operators $P_0$, $P_j$, $J_j$, $K_j$, $j=1,2,3$ exist,
which generate the ten one-parameter unitary subgroups 
$$
\{e^{iP_0t}\}\, ,\quad\{e^{-i{P}_j a_j},\,a\in{\RR}\}\,,\quad
\{e^{-i{J}_j \theta_j},\,\theta_j\in{\RR}\}\, ,\quad\{e^{-i{K}_j \varphi(u_j)},\,u_j\in{\RR}\}\eqno(2)
$$
of ${\mathcal U}({\mathcal H})$, which  represent the one-parameter
sub-groups $\tilde{\mathcal T}_0$, $\tilde{\mathcal T}_j,\tilde{\mathcal R}_j$, $\tilde{\mathcal B}_j$ according to the 
representation $U$.
The mathematical structural properties of $\tilde{\mathcal P}_+^\uparrow$ as a Lie group imply that
these generators satisfy the following commutation relations \cite{Wigner1},\cite{Foldy}.
\vskip.5pc\noindent
(i)\hskip4.3mm $[{P}_j,{P}_k]=\nop$,\hskip15mm (ii) $[{J}_j,{P}_k]=i{\hat\epsilon}_{jkl}{P}_l$,
\hskip8mm (iii) $[{ J}_j,{ J}_k]=i{\hat\epsilon}_{jkl}{ J}_l$,\par\noindent
(iv)\hskip2.3mm $[{ J}_j,{ K}_k]
=i{\hat\epsilon}_{jkl}{ K}_l$,\quad\hskip1.8mm (v)\hskip1mm $[{ K}_j,{ K}_k]=-i\hat\epsilon_{j,k,l}J_l$,
\;\;(vi) $[{ K}_j,{ P}_k]=i\delta_{jk}P_0$,\hfill{(3)}
\par\noindent
(vii) $[P_j,P_0]=\nop$,\hskip15.2mm (viii) $[J_j,P_0]=\nop$,\hskip14.8mm (ix) $[K_j,P_0]=iP_j$,
\vskip.4pc\noindent
where ${\hat\epsilon}_{jkl}$ is the Levi-Civita symbol ${\epsilon}_{jkl}$
restricted by the condition $j\neq l\neq k$.
\vskip.5pc
Given a reference frame $\Sigma$ in  $\mathcal F$ and $g\in\mathcal P$, we introduce  the mapping 
$\textsf g:\RR^4\to\RR^4$ such that if 
$\underline x=(t,x_1,x_2,x_3)\equiv(x_0,{\bf x})$ is the vector of the time-space coordinates of an event with respect to $\Sigma$, 
then $\textsf g(\underline x)$ is the vector of the time-space coordinates of that event with respect to the frame $\Sigma_g$
related to $\Sigma$ just by $g$.
Once introduced the four-operator $\underline P=(P_0,P_1,P_2,P_3)\equiv(P_0,{\bf P})$, from (3)  it can be deduced that the following
statement holds for all $\tilde g\in\tilde{\mathcal L}_+^\uparrow$.
$$
U_{\tilde g}\underline P U_{\tilde g}^{-1}=\textsf g(\underline P), \hbox{ where } g=\textsf h(\tilde g).\eqno(4)
$$
\vskip.5pc
The mathematical structural properties of the {\sl full} Poincar\'e group $\mathcal P$ allow to extend 
(3) to include $\SREV,\TREV$ according to the following statements \cite{Wigner1},\cite{Foldy}:
\vskip.4pc\noindent
If $\SREV$ {\sl is unitary}, then its phase factor can be chosen so that $\SREV^2=\Id$, and
\par\noindent
$[\SREV,P_0]=\nop$,\quad $\SREV P_j=-P_j\SREV$,\quad $[\SREV,J_k]=\nop$,\quad $\SREV K_j=-K_j\SREV$;\quad
;\hfill{(5)}
\vskip.4pc\noindent
If $\SREV$ is anti-unitary, then $\SREV^2=c\Id$, so that $\SREV^{-1}=c\SREV$, where $c=1$ or $c=-1$, and
\par\noindent
$\SREV P_0=-P_0\SREV$,\quad $[\SREV, P_j]=\nop$,\quad $\SREV J_k=-J_k\SREV$,\quad $\SREV K_j=K_j\SREV$,\hfill{(6)}
\vskip.5pc\noindent
 If $\TREV$ is unitary, then its phase factor can be chosen so that \quad$\TREV^2=\Id$, and
\par\noindent
$\TREV P_0=-P_0\TREV$,\quad $[\TREV,P_j]=\nop$,\quad $[\TREV,J_k]=\nop$,\quad $\TREV K_j=-K_j\TREV$;\hfill{(7)}
\vskip.5pc\noindent
If $\TREV$ is anti-unitary, then $\TREV^2=c\Id$, so that $\TREV^{-1}=c\TREV$, either $c=1$ or $c=-1$, and
\par\noindent
$\TREV P_0=P_0\TREV$,\quad $\TREV P_j=-P_j\TREV$,\quad $\TREV J_k=-J_k\TREV$,\quad $\TREV K_j=K_j\TREV$,\hfill{(8)}
\vskip.5pc\noindent
$\SREV\TREV=\omega\TREV\SREV$, \quad with $\omega\in\CC$ and $\vert\omega\vert=1$\hfill{(9)}
\vskip.5pc
The {\sl helicity} operator is defined by $\hat\lambda=\frac{{\bf J}\cdot{\bf P}}{P}$, where $P=\sqrt{P_1^2+P_2^2+P_3^2}$.
The following relation
$$\SREV\hat\lambda\SREV^{-1}=-\hat\lambda \eqno(10)$$
is implied by (5) but also by (6). Therefore it holds independently of the unitary or anti-unitary character of $\SREV$.
\par
By making use of (3)-(8) it can be proved that the following relations hold.
$$
[V,P_0^2-{\bf P}^2]=\nop,\quad[V,W_0^2-{\bf W}^2]=\nop,\quad
\hbox{for all }V\in U(\tilde{\mathcal P}_+^\uparrow)\cup\{\TREV,\SREV\}\eqno(11)
$$
where $W_0={\bf J}\cdot{\bf P}$ and ${\bf W}=P_0{\bf J}+{\bf P}\land{\bf K}$ form the
{\sl Pauli-Luba\'nski four-operator} $(W_0,{\bf W})$.

%%___________________________________________________________________________________________________  3.3
\subsection{ Massless elementary free particles}
An isolated system is said to be {\sl localizable} if its quantum theory is endowed 
with a unique {\sl position observable}, that is to say with a {\sl unique} tern $(Q_1,Q_2,Q_3)\equiv{\bf Q}$ of self-adjoint operators,
whose components $Q_j$ are called {\sl coordinate} operators, characterized by the following conditions
\footnote{
The commutativity condition $[Q_j,Q_k]=\nop$ establishes the possibility of performing a measurement
that yields all three values of the position coordinates.
The {\sl nonexistence} of commutative position operators in certain circumstances \cite{Wightman} 
prompted to search for {\sl non-commutative}  \cite{Kalnay},\cite{JadzickAl},\cite{Bacry1},\cite{Bacry2} or 
{\sl unsharp} position operators (see \cite{Schroeck} and references therein),
In the present work we are interested in the commutative concept of position only.}.
\begin{itemize}
\item[({\it Q}.1)]
$[Q_j,Q_k]=\nop$, for all $j,k=1,2,3$.
\item[]
This condition establishes that a measurement of position yields
all three values of the coordinates of the same specimen of the system.
\item[({\it Q}.2)]
For every $g\in\mathcal P$,
the tern $(Q_1,Q_2,Q_3)\equiv{\bf Q}$ and the transformed position operators 
$S_g[{\bf Q}]\equiv (S_g[{\bf Q}_1],S_g[{\bf Q}_2],S_g[{\bf Q}_3])$ satisfy
the specific  relations implied by the transformation properties of position with respect to $g$.
\end{itemize}
 According to ({\it Q},2), the following specific relations hold.
\begin{itemize}
\item[]\quad
$S_\trev[{\bf Q}]={\bf Q}$ and $S_\srev[{\bf Q}]=-{\bf Q}$, i.e. $\TREV{\bf Q}={\bf Q}\TREV$ and 
$\SREV{\bf Q}=-{\bf Q}\SREV$,\hfill{(12.i)}
\item[]\quad
$S_g[{\bf Q}]=U_g{\bf Q}U_g^{-1}={\textsf g}({\bf Q})$ for every $g\in{\mathcal E}$, which imply
\item[]\quad 
$[Q_k,P_j]=i\delta_{jk}$ and $[J_j,Q_k]=i\hat\epsilon_{jkl}Q_l.$\hfill{(12.ii)}
\end{itemize}
A localizable isolated system will be also called {\sl free particle}.
Following a customary procedure, a free particle will be qualified as {\sl elementary} if the system of operators
$\{U(\tilde{\mathcal P}_+^\uparrow),\SREV,\TREV; {\bf Q}\}$ is irreducible.
\par
For an elementary free particle, the transformer tern $(U,\SREV,\TREV)$ of its quantum theory 
must be {\sl irreducible}. Let us explain why.
If $(U,\SREV,\TREV)$ were reducible, then a unitary operator $V$ would exist such that 
$[\SREV,V]=[\TREV,V]=[V,U_{\tilde g}]=\nop$ for all $\tilde g\in\tilde{\mathcal P}_+^\uparrow$,
but $[V,Q_j]\neq\nop$ for some $j$ $-$ if $[V,Q_k]=\nop$ held for all $k$, then
$\{U(\tilde{\mathcal P}_+^\uparrow),\SREV,\TREV; {\bf Q}\}$ would be reducible, and this is not possible for elementary particles.
Hence, if we define
$\hat Q_k=VQ_kV^{-1}$, then $\hat{\bf Q}\neq {\bf Q}$, while $VU_gV^{-1}=U_g$ for all $g\in\mathcal P$.
The mathematical relations between the operators $\hat{\bf Q}=V{\bf Q}V^{-1}$ and each operator 
$V BV^{-1}$, where $B$ belongs to the tern $(U,\SREV,\TREV)$ must be the same as the mathematical relations between
$\bf Q$ and that $B$, because $\{U(\tilde{\mathcal P}_+^\uparrow),\SREV,\TREV; {\bf Q}\}$ and
$V\{U(\tilde{\mathcal P}_+^\uparrow),\SREV,\TREV; {\bf Q}\}V^{-1}$ are unitarily isomorphic.
Then the tern $\hat{\bf Q}$ satisfies ({\it Q}.2), because $\bf Q$ does, and therefore
$\hat{\bf Q}$ would be a position operator in all respects.
Thus, for the same elementary particle two different position operators would exist, in contradiction with the required uniqueness.
\vskip.5pc
Since the representation $U$ of $\tilde{\mathcal P}_+^\uparrow$ can be reconstructed from its generators according 
to (2), a transformer tern will be irreducible if and only if the operators $P_0,P_j, J_j,K_j, \SREV,\TREV, j=1,2,3$ 
form an irreducible system; therefore, by a straightforward application of Schur lemma, from (11) we imply that
the quantum theory of an elementary free particle is
characterized by two real numbers $\eta$, $\varpi$  such that 
$$
P_0^2-{\bf P}^2=\eta\Id,\qquad 
W^2\equiv W_0^2 -(W_1^2+W_2^2+W_3^2)=\varpi\Id\,.\eqno(13)
$$
In the present work we investigate the theories of those isolated systems for which $\eta=0$, i.e. $P=\vert P_0\vert$, 
called {\sl massless} systems.
%____________________________________________________________________________________________________  4
\section{Identification of theories of massless systems}
Because of the necessary irreducibility of the transformer tern of every quantum theory of a massless elementary free particle,
the first step for identifying these theories is to identify the class of all irreducible transformer terns with $\eta=0$. 
Since a reducible tern is decomposable in terms of irreducible terns, in so doing we identify also the class of all terns
any theory of massless isolated system is built up.
\par
Now, the irreducible components of the representation $U$ in such a transformer tern are
irreducible representations of $\tilde{\mathcal P}_+^\uparrow$ with $\eta=0$, which are well known.
Each of them is characterized by a non-negative value of a parameter $r_0$ related to the Pauli-Lubanski constant $\varpi$ in (13).
If $r_0>0$, then the helicity operator turns out to have unbounded spectrum \cite{Wigner1},\cite{Wightman}, therefore the
set of the possible values of the helicity $\hat\lambda$ is unbounded. 
This feature has never been observed in Nature;  the present investigation is restricted to the empirically meaningful theories,
that is to say to theories  with bounded helicity. Accordingly, we shall identify the class
${\mathcal I}$ of the irreducible terns whose representation has all irreducible components with $\eta=0$ and $r_0=0$.
To this aim, we shall make use of the following classification of these terns in terms of spectral properties of the representation $U$.

%______________________________________________________________________________________________________________ 4.1.1
\subsection{Three kinds of theories}
The irreducibility condition for the transformer tern $(U,\SREV,\TREV)$ imposes strong constraints to the
spectrum
$\sigma(\underline P)$ of the four-operator $\underline P$.
Namely there are three mutually exclusive 
possibilities for $\sigma(\underline P)$, and each possibility turns out to be related to
the unitary or anti-unitary character of  the space inversion and time reversal operators $\SREV$ and $\TREV$.
A clear formulation of this fact requires some notion of spectral theory.
\par
Let $E^{(\alpha)}:{\mathcal B}(\RR)\to\Pi({\mathcal H})$ be the spectral measure 
\cite{ReedAl} of each generator $P_\alpha$, $\alpha=0,1,2,3$.
Since the operators $P_\alpha$ commute with each other, a common spectral measure $E:{\mathcal B}(\RR^4)\to\Pi({\mathcal H})$
is generated by defining it for  hyperrectangles $\Delta=\Delta_0\times \Delta_1\times\Delta_2\times\Delta_3$, where $\Delta_\alpha=(a_\alpha,b_\alpha]$,
as $E(\Delta)=E^{(0)}(\Delta_0)E^{(1)}(\Delta_1)E^{(2)}(\Delta_2)E^{(3)}(\Delta_3)$; then
$$
\underline P =\int \underline p\,dE{\underline p}\,,\eqno(14)
$$
with $dE_{\underline p}=E\left((p_0,p_0+dp_0]\times(p_1,p_1+dp_1]\times (p_2,p_2+dp_2]\times (p_3,p_3+dp_3] \right)$.
\par
Once  introduced the {\sl four-operator} $\underline P=(P_0,P_1,P_2,P_3)\equiv(P_0,{\bf P})$,
the {\sl spectrum} of $\underbar P$ can be defined as the following subset of $\RR^4$.
$$
\sigma({\underbar P})=\{\underbar p=(p_0,{\bf p})\in\RR^4\mid
E(\Delta_{\underbar p})\neq\nop\hbox{ for every neighborough }\Delta_{\underbar p}\hbox{ of }
\underbar p\}\,.\eqno(15)
$$
\vskip.5pc\noindent
{\bf Proposition 4.1.}
{\sl  Let $U:\tilde{\mathcal P}_+^\uparrow\to {\mathcal U}({\mathcal H})$ be a continuous unitary  representation, and let
$\SREV,\TREV\in{\mathcal V}({\mathcal H})$ satisfy (5)-(9).
If $(U;\SREV,\TREV)$ is irreducible with $\eta=0$, then
there are only the following mutually exclusive possibilities for the spectra $\sigma(\underbar P)$ and $\sigma(P_0)$.
\vskip.3pc\noindent
({\bf u})\quad $\sigma(\underbar P)=S^{+}_0$ and $\sigma(P_0)=(0,\infty)$, with $\SREV$ unitary and $\TREV$ anti-unitary,
\vskip.3pc\noindent
({\bf d})\quad $\sigma(\underbar P)=S^{-}_0$ and $\sigma(P_0)=(-\infty,0)$,  with $\SREV$ unitary and $\TREV$ anti-unitary,
\vskip.3pc\noindent
({\bf s})\quad $\sigma(\underbar P)=S^{+}_0\cup S^{-}_0$ and $\sigma(P_0)=\RR\setminus\{0\}$, 
 with $\SREV$ anti- unitary  or $\TREV$ unitary,
\vskip.3pc\noindent
where $S^{+}_0=\{\underbar p\mid p_0^2-{\bf p}^2=0,\,p_0>0\}$ and 
$S^{-}_0=\{\underbar p\mid p_0^2-{\bf p}^2=0^2,\,p_0<0\}$ are the positive and the negative hypercones, or {\sl zero-mass orbits},
 in $\RR^4$.
}
\vskip.5pc\noindent
A proof can be found in \cite{Dice}  and references therein.
%%%%%%%%%%%%%%_________________________________________________________________________________________   4.1.2
\subsubsection{Implications of the irreducibility of $U$}
If just the representation $U$ of a transformer tern $(U;\SREV,\TREV)$ is irreducible, then the whole tern is irreducible, of course. 
In such a case, however, only the cases ({\bf u}) or ({\bf d}) of Prop. 4.1 can occur, because of the following proposition.
\vskip.5pc\noindent
{\bf Proposition 4.2.}
{\sl Once defined the projection operators $E^{+}=\int_{S^+_0}dE_{\underline p}\equiv\int_0^\infty p_0dE_{p_0}^{(0)}$
and $E^{-}=\int_{S^-_0}dE_{\underline p}\equiv\int_{-\infty}^{0} p_0dE_{p_0}^{(0)}$, with
ranges ${\mathcal H}^+=E^+{\mathcal H}$ and ${\mathcal H}^-=E^-{\mathcal H}$, respectively,
the relation $[E^+,U_{\tilde g}]=\nop$ and $[E^-,U_{\tilde g}]=\nop$ hold for all $\tilde g\in\tilde{\mathcal P}_+^\uparrow$ .}
\vskip.4pc\noindent
{\bf Proof.}
Since $E^+=\int_{S^+_0}dE_{\underline p}\equiv\chi_{S^+_0}(\underline P)$,
where $\chi_{S^+_0}$ is the characteristic functional of $S^+_0$, the relations (1.i),(1.vii) imply that
$E^+$ commutes with $P_0$ and with all $P_j$, and therefore ${\mathcal H}^+$ is left invariant by 
$U_{\tilde g}$, if $\tilde g\in{\mathcal T}_0\cup{\mathcal T}_1\cup{\mathcal T}_2\cup{\mathcal T}_3$. 
Now we show that ${\mathcal H}^+$ is left invariant by
$U_{\tilde g}$, for every $\tilde g\in\tilde{\mathcal L}_+^\uparrow$ too, and hence for all $\tilde g\in\tilde{\mathcal P}_+^\uparrow$.
Relation (4) implies that $U_{\tilde g}E(\Delta)U_{\tilde g}^{-1}=E(\textsf g^{-1}(\Delta)$ holds,
where $g=\textsf h(\tilde g)$; therefore,
if $\psi\in{\mathcal H}^+$, then for every $\tilde g\in\tilde{\mathcal L}_+^\uparrow$ we have
$U_{\tilde g}\psi=U_{\tilde g}\int_{S^+_0}dE_{\underline p}\psi=
\int_{{S^+_0}}dU_{\tilde g}E_{\underline p}{U_{\tilde g}}^{-1}(U_{\tilde g}\psi)
=\int_{S^+_0}dE_{{\textsf g}^{-1}({\underline p})}(U_{\tilde g}\psi)$.
The last integral is a vector of ${\mathcal H}^+=E^+\mathcal H$, because
 ${\underline p}'={\textsf g}^{-1}(\underline p)\in S_0^+$ if $\underline p\in S_0^+$ for $\tilde g\in\tilde {\mathcal L}_+^\uparrow$.
The proof of $[E^-,U_{\tilde g}]=\nop$ is quite analogous,\hfill{$\bullet$}
\vskip.5pc\noindent
Hence, according to Prop. 4.2, if $U$ is irreducible then either the case $E^+=\Id$, 
$E^-=\nop$ or the case $E^+=\nop$ and $E^-=\Id$ must occur.
If the first case occurs, then the relation
$\underline P=E^+\underline P=\chi_{S_0^+}(\underline P)\underline P$ holds, which implies 
$\sigma(\underline P)=\{\underline p\in\RR^4,\;\chi_{S_0^+}(\underline p)=\underline p, \,\}$,
 i.e. $\sigma(\underline P)=S_0^+$.
Analogously, in the alternative case where $E^-=\Id$ occurs, $\sigma(\underline P)=S_0^-$ is obtained.
\vskip.5pc
\par
According to the classification operated by Prop. 4.1, the class $\mathcal I$ is the union ${\mathcal I}={\mathcal I}({\bf u})\cup{\mathcal I}({\bf d})\cup{\mathcal I}({\bf s})$
of three non-overlapping subclasses; the sublclass ${\mathcal I}({\bf u})$ (resp.,  ${\mathcal I}({\bf d})$, ${\mathcal I}({\bf s})$) is formed by 
those irreducible terns in $\mathcal I$ for which $\sigma(\underline P)=S_0^+$ (resp., $S_0^-$, $S_0^-\cup S_0^+$). So, there 
can be three kinds of theories, according to the sub-class ${\mathcal I}({\bf u})$,
 ${\mathcal I}({\bf d})$ or ${\mathcal I}({\bf s})$ the transformer tern of the theory belongs to.
The representation $U$ in a tern of $\mathcal I$ is the direct sum or integral of irreducible representations $U^\alpha$.
In section 5 we determine the features of the transformer terns of ${\mathcal I}({\bf u})$ and ${\mathcal I}({\bf d})$. The terns of 
${\mathcal I}({\bf s})$ will be investigated in section 6.
%____________________________________________________________________________________________________   5
\section{Theories with terns in ${\mathcal I}({\bf u})$ and  ${\mathcal I}({\bf d})$.}
The representation $U$ of a transformer tern $(U;\SREV,\TREV)$ in the class $\mathcal I({\bf u})$ or $\mathcal I({\bf d})$ can be 
reducible or not, though the tern is irreducible. 
In subsection 5.1 we determine conditions to be satisfied by an irreducible representation in order to give rise 
to a theory based on ${\mathcal I}({\bf u})$.
In section 5.2 these conditions  are proved to apply to the theories with $U$ reducible too.
In section 5.3 this treatment is extended to  ${\mathcal I}({\bf d})$.
%____________________________________________________________________________________________________ 5.1
\subsection{Theories based on terns with irreducible $U$}
According to Prop. 4.2,
a transformer tern with $U$ irreducible cannot belong to ${\mathcal I}({\bf s})$.
If $r_0=0$, then within unitary isomorphisms each irreducible representation 
of ${\mathcal P}_+^\uparrow$ with $\sigma(\underline P)=S_0^+$ or $\sigma(\underline P)=S_0^-$
turns out to be completely identified by a number $m\in\ZZ$.
Representations with different values of this paramenter are unitarily inequivalent.
{\it Modulo} unitary isomorphisms, the Hilbert space of the representation is ${\mathcal H}=L_2(\RR^3, d\nu)$,
i.e. the space of the complex functions on $\RR^3$ square integrable with respect to the invariant measure
$d\nu=\frac{dp_1\,dp_2\,dp_3}{p_0}$, with $p_0=\sqrt{{\bf p}^2}$.
\par
The generators of the representation with $\sigma(\underline P)=S_0^+$ and a given $m$ are the following operators.
\vskip.4pc\noindent
$(P_j\psi)({\bf p})=p_j\psi({\bf p})$, \quad$P_0\psi({\bf p})=p_0\psi({\bf p})$\hfill{(16)}\vskip.4pc\noindent
$J_j=J_j^{(0)}+\textsf j_j$,\quad $K_j=K_j^{(0)}+\textsf k_j$,\hfill{(17)}
\vskip.4pc\noindent\noindent
where $J_j^{(0)}=-i\left(p_k\frac{\partial}{\partial p_l}-p_l\frac{\partial}{\partial p_k}\right)$, $(j,k,l)$ 
 being  cyclic, $K_j^{(0)}=ip_0\frac{\partial}{\partial p_j}$, and\vskip.4pc\noindent
$\textsf j_1=\frac{m}{2}\frac{p_1p_0}{p_1^2+p_2^2}$,\;\;
$\textsf j_2=\frac{m}{2}\frac{p_2p_0}{p_1^2+p_2^2}$,\;\;
$\textsf j_3=0$,\;\;
$\textsf k_1=-\frac{m}{2}\frac{p_2p_3}{p_1^2+p_2^2}$,\;\;
$\textsf k_2=\frac{m}{2}\frac{p_3p_1}{p_1^2+p_2^2}$,\;\;
$\textsf k_3=0$.\hfill{(18)}
\vskip.5pc
If $\sigma(\underline P)=S_0^-$ the generators are
\vskip.4pc\noindent
$(P_j\psi)({\bf p})=p_j\psi({\bf p})$, \quad$P_0\psi({\bf p})=-p_0\psi({\bf p})$\hfill{(19)}\vskip.4pc\noindent
$J_j=J_j^{(0)}+\textsf j_j$,\quad $K_j=-(K_j^{(0)}+\textsf k_j)$.\hfill{(20)}
\vskip.4pc\noindent
A simple computation shows that (16)-(20)
imply $\hat\lambda=\frac{m}{2}\Id$, so that any representation in the class we are studying turns out to be 
identified by the value of the helicity. 
\vskip.4pc
Possible theories are obtained by identifying 
the space inversion operator $\SREV$ and the time reversal 
operator $\TREV$ consistent with (16),(17) to form a tern of ${\mathcal I}({\bf u})$, or consistent with (19),(20) 
to form a tern of ${\mathcal I}({\bf d})$.
According to Prop. 4.1 the operator $\SREV$ must be unitary and $\TREV$ must be anti-unitary. 
Relations (5), 
implies $\SREV\hat\lambda\SREV^{-1} =-\hat\lambda$, i.e., $\frac{m}{2}=-\frac{m}{2}$.
Therefore, all terns in ${\mathcal I}({\bf u})$ or ${\mathcal I}({\bf d})$ with $U$ irreducible have $m=0$ and hence zero helicity.
\par
To explicitly identify the unitary operator$\SREV$,  let us define $\hat S=\Upsilon\SREV$. Then
$[\SREV,P_0]=\nop$ and $\SREV P_j=-P_j\SREV$ in
(5) imply $[\hat S, P_j]=\nop$; hence $\hat S\psi({\bf p})=s({\bf p})\psi({\bf p})$  where $s$ is a complex
function of $\bf p$ because $(P_1,P_2,P_3)$ in (16) form a complete system of operators in $L_2(\RR^3,d\nu)$.
Taking into account (16)-(20), the last equation in (5) becomes
$K_j^{(0)}\Upsilon s=-\Upsilon s K_j^{(0)}$, which implies that $s$ is a 
constant function, hence we can set $s=1$; so that $\SREV=\Upsilon$.
\par
To explicitly find $\TREV$ we introduce the unitary operator $\hat T=\mathcal K\Upsilon\TREV$, where 
$\mathcal K$ is the anti-unitary complex conjugation operator defined by $\mathcal K\psi({\bf p})=\overline{\psi({\bf p})}$ 
and $\Upsilon$ is the unitary parity operator defined by $\Upsilon\psi({\bf p})=\psi(-{\bf p})$.
By applying  (8) to the operators of (16)-(20) we find 
$[\hat T,P_0]=[\hat T,P_j]=[\hat T,J_j]=[\hat T,K_j]=\nop$ for all $j$. 
Since $U$ is irreducible, the set of its generators is irreducible too, so that $\hat T$ is a  
constant operator $\hat T=e^{i\theta_0}\Id$; thus we
can choose the phase factor so that $\TREV=\mathcal K\Upsilon$.
\par
Thus, modulo unitary isomorphisms, each sub-class ${\mathcal I}({\bf u})$ or ${\mathcal I}({\bf d})$ contains only one irreducible 
tranformer tern with $r_0=0$ and $U$ irreducible, identified
\vskip.5pc\noindent
by  (16) and $J_j=J_j^{(0)}$, $K_j=K_j^{(0)}$, $\SREV=\Upsilon$, $\TREV={\mathcal K}\Upsilon$,
in ${\mathcal I}({\bf u})$, and
\hfill{(21)}
\vskip.5pc\noindent
by
(19) and
$J_j=J_j^{(0)}$, $K_j=-K_j^{(0)}$, $\SREV=\Upsilon$, $\TREV={\mathcal K}\Upsilon$, in ${\mathcal I}({\bf d})$.
\hfill{(22)\par

%_________________________________________________________________________________________________  5.2
\subsection{Constraints for terns of ${\mathcal I}({\bf u})$ or ${\mathcal I}({\bf d})$ with reducible $U$}
Since the irreducible representations of ${\mathcal P}_+^\uparrow$ with $r_0=0$ and $\eta=0$ form a
{\sl discrete} set, the representation $U$ in a tern of  ${\mathcal I}({\bf u})$ (resp.,  ${\mathcal I}({\bf d})$)
must be the direct sum  $U=\oplus_\alpha U^\alpha$ of representations identified by (16)-(18) (resp., (19)-(20)), so that
the Hilbert space of the theory 
is the direct sum ${\mathcal H}=\oplus_\alpha {\mathcal H}_\alpha$, where ${\mathcal H}_\alpha$ is the
Hilbert space of  the irreducible component $U^\alpha$.
In this section we prove that also in this case the helicity must be zero.
\vskip.5pc\noindent
{\bf Proposition 5.1.}
{\sl The helicity of a theory with irreducible tern in ${\mathcal I}({\bf u})\cup{\mathcal I}({\bf d})$ is zero,
independently of the reducibility of $U$.}
\vskip.4pc\noindent
{\bf Proof.}
Every irreducible component $U^\alpha$ of $U$ has $r_0=0$, and hence it is characterized by a specific value $m_\alpha$ of the paramenter
$m$ in (16)-(20). We prove that $m_{\alpha}=0$, for every $\alpha$.
\par
Let us suppose that $m_{\alpha_0}\equiv m\neq 0$. Given any vector 
$\psi\in{\mathcal H}_{\alpha_0}$ with $\Vert\psi\Vert=1$,
the vector $\tilde\psi=\SREV\psi$ must be not null because $\SREV$ is unitary. Being $\tilde\psi$ in 
${\mathcal H}=\oplus_\alpha {\mathcal H}_\alpha$, it can be decomposed as
$\tilde\psi=c_0\varphi_{\alpha_0}+\sum_{\alpha\neq\alpha_0}c_\alpha\varphi_\alpha$, with $\varphi_\alpha\in{\mathcal H}_\alpha$
and $c_\alpha\in\CC$. According to (10)
$\SREV\hat\lambda\SREV^{-1}=-\hat\lambda$ holds, which implies\par\noindent
$\hat\lambda\tilde\psi=c_0\hat\lambda\varphi_{\alpha_0}+\sum_{\alpha\neq\alpha_0}c_\alpha\hat\lambda\varphi_\alpha=
\frac{m}{2}c_0\varphi_{\alpha_0}+\sum_{\alpha\neq\alpha_0}(\frac{m_{\alpha}}{2})c_\alpha\varphi_\alpha$
\par
$=
\SREV\SREV^{-1}\hat\lambda\SREV\psi=\SREV(-\hat\lambda)\psi=-\frac{m}{2}\SREV\psi=-\frac{m}{2}\tilde\psi
=-\frac{m}{2}c_0\varphi_{\alpha_0}-\sum_{\alpha\neq\alpha_0}\frac{m}{2}c_\alpha\varphi_\alpha$.
\par\noindent
Hence $ \frac{m}{2}c_0\varphi_{\alpha_0}+\sum_{\alpha\neq\alpha_0}(\frac{m_{\alpha}}{2})c_\alpha\varphi_\alpha=
-\frac{m}{2}c_0\varphi_{\alpha_0}-\sum_{\alpha\neq\alpha_0}\frac{m}{2}c_\alpha\varphi_\alpha$;
since $\langle\varphi_\alpha\mid\varphi_\beta\rangle=\delta_{\alpha\beta}$, we have to conclude that
$c_0=0$ and if $c_\alpha\neq 0$ then $m_\alpha=-m$.
Therefore, if there are component representations $U^{\alpha_0}$ with $m_{\alpha_0}=m\neq 0$, then there must be also
component representations with $m_\alpha=-m$. Let $\{U^{\alpha_j^+}\}$ and $\{U^{\alpha_k^-}\}$ be 
the set of all irreducible components with $m_{\alpha_j^+}=m$ and $m_{\alpha_k^-}=-m$, and let us define the representations
$U^{(+)}=\oplus_jU^{\alpha_j^+}$ and $U^{(-)}=\oplus_kU^{\alpha_k^-}$, which act of the Hilbert spaces
 and ${\mathcal H^{(+)}}=\oplus_{\alpha_j^+}{\mathcal H}_{\alpha_j^+}$ and 
${\mathcal H}^{(-)}=\oplus_{\alpha_k^-}{\mathcal H}_{\alpha_k^-}$ respctively.
\par
Every vector $\psi\in\mathcal H$  can be decomposed as $\psi=\psi^++\psi^-+ \psi_0$, where 
$\psi^+\in{\mathcal H}^{(+)}$, $\psi^-\in{\mathcal H}^{(-)}$ and 
$\psi_0\in{\mathcal H}^{(0)}=\oplus_{\alpha_j^+\neq\alpha\neq\alpha_k^-}{\mathcal H}_\alpha$.
Of course, ${\mathcal H}^{(0)}\bot\left({\mathcal H}^{(+)}\oplus{\mathcal H}^-\right)$.
Such a decomposition induces a matrix representation where the vector $\psi$ is
represented by a column vector $\psi\equiv\left[\begin{matrix}\psi^+\cr \psi^-\cr\psi_0\end{matrix}\right]$ and
any linear or anti-linear operator $A$ is represented by a matrix
$A\equiv\left[\begin{matrix}A_{++}&A_{+-}&A_{+0}\cr A_{-+}&A_{--}&A_{-0}\cr
A_{0+}&A_{0-}&A_{00}\end{matrix}\right]$. Accordingly,  in the case that the tern belongs to ${\mathcal I}({\bf u})$,
the generators are represented by\par\noindent
$P_0=\left[\begin{matrix}p_0&0&0\cr0&p_0&0\cr0&0&p_0\end{matrix}\right]$,\quad
$P_j=\left[\begin{matrix}p_j&0&0\cr0&p_j&0\cr0&0&p_j\end{matrix}\right]$,\par\noindent
$J_j=\left[\begin{matrix}(J_j^{(0)}+\textsf j_j)&0&0\cr0&(J_j^{(0)}-
\textsf j_j)&0\cr0&0&J_{j00}\end{matrix}\right]$,\quad
$K_j=\left[\begin{matrix}(K_j^{(0)}+\textsf k_j)&0&0\cr0&(K_j^{(0)}-
\textsf k_j)&0\cr0&0&K_{j00}\end{matrix}\right]$.
\par\noindent
Now we show that no space inversion operator exists if $m\neq 0$. Let us introduce the unitary operator 
$\hat S=\Upsilon\SREV$.
By imposing (5) for the matrices  $P_j$, by the completeness of $p_j$ in $L_2(\RR^3,d\nu)$ we imply that
$\hat S=\left[\begin{matrix}0&S_1({\bf p})&0\cr S_2({\bf p})&0&0\cr0&0&
S_{00}({\bf p})\end{matrix}\right]$, where $S_1$ and $S_2$ are functions of $\bf p$.
\par\noindent
 For the matrices $K_j$, $j=1,2$ condition (5) implies
$2iS_1({\bf p})p_0\frac{\partial}{\partial p_j}=2\textsf k_j-ip_0\frac{\partial S_1({\bf p})}{\partial p_j}$, 
$2iS_2({\bf p})p_0\frac{\partial}{\partial p_j}=-2\textsf k_j+ip_0\frac{\partial S_2({\bf p})}{\partial p_j}$
which cannot be satisfied unless $S_1=S_2=\nop$.
\par\noindent
The proof for terns in ${\mathcal I}({\bf d})$ is quite analogous.\hfill{$\bullet$}
\vskip.5pc\noindent
According to this result, in a theory based on a transformer tern in ${\mathcal I}({\bf u})\cup {\mathcal I}({\bf d})$ with $U$ reducible,
all irreducible components $U^\alpha$ must be identical to each other, up unitary isomorphism, and with $m_\alpha=0$.
\vskip.5pc\noindent
{\bf Example 5.1.}
Let us consider
the Hilbert space
${\mathcal H}={\mathcal H}^{(1)}\oplus {\mathcal H}^{(2)}$, with ${\mathcal H}^{(1)}={\mathcal H}^{(2)}= L_2(\RR^3,\CC^{2s+1},d\nu)$;
let every vector $\psi=\psi_1+\psi_2$ in $\mathcal H$, with $\psi_1,\psi_2\in  L_2(\RR^3,\CC^{2s+1},d\nu)$
be represented
as the column vector $\psi\equiv \left[\begin{array}{c}\psi_1\cr \psi_2\end{array}\right]$.
The operators represented by the matrices
$$
P_0=\left[\begin{array}{cc}p_0& 0 \cr 0 & p_0\end{array}\right]\,,\;
P_j=\left[\begin{array}{cc}p_j& 0 \cr 0 & p_j\end{array}\right]\,,\;
J_k=\left[\begin{array}{cc}J_k^{(0)}& 0 \cr 0 &J_k^{(0)}\end{array}\right]\,,\;
K_j=\left[\begin{array}{cc}K_j^{(0)}& 0 \cr 0 &K_j^{(0)}\end{array}\right]\,,
$$
satisfy (3).  Hence,
a {\sl reducible}  representation $U:\tilde{\mathcal P}_+^\uparrow\to
{\mathcal H}^{(1)}\oplus {\mathcal H}^{(2)}$,
where ${\mathcal H}^{(1)}={\mathcal H}^{(2)}L_2(\RR^3,\CC^{2s+1},d\nu)$, 
is determined. The operators
$$\SREV=\Upsilon\left[\begin{array}{cc}0&1\cr 1&0\end{array}\right]\,,\quad
 \TREV=\Upsilon{\mathcal K}\left[\begin{array}{cc}0& 1 \cr -1 & 0\end{array}\right]
\eqno(23)
$$
satisfy conditions (5),(8),(9). Therefore, an irreducible transformer tern of ${\mathcal I}({\bf u})$ is identified. Moreover,
if $A$ is any self-adjoint operator of 
${\mathcal H}=L_2(\RR^3,\CC^{2s+1},d\nu)\oplus L_2(\RR^3,\CC^{2s+1},d\nu)$, then
the conditions $[A,P_0]=[A,P_j]=[A,J_k]=[A,K_j]=[A,\SREV]=[A,\TREV]=\nop$ imply 
$A=a\Id$, and therefore the tern is irreducible.
\vskip.5pc\noindent
{\bf Example 5.2.} An irreducible tern of ${\mathcal I}({\bf d})$ with $U$ reducible 
can be obtained from Example 5.1, by changing only $P_0$ and $K_j$ into
$$
P_0=\left[\begin{array}{cc}-p_0& 0 \cr 0 &- p_0\end{array}\right]\,,\quad
K_j=\left[\begin{array}{cc}-K_j^{(0)}& 0 \cr 0 &-K_j^{(0)}\end{array}\right]\,,\eqno(24)
$$
{\bf Remark 5.1.}
The necessity for theories based on transformer terns with reducible $U$ arose
from the empirical evidence that specific massless particles, namely photons,
occur with positive and negative helicities \cite{RamanAl}.
Then theories were taken into account based on irreducible terns
whose representation $U$ is the direct sum of two irreducible representaions of $\tilde{\mathcal P}_+^\uparrow$ \cite{Amrein},
\cite{Weinberg1},\cite{Jordan1}, characterized by opposite non-zero values of the helicity, but with $\SREV$  unitary and $\TREV$  anti-unitary.
Prop. 4.1 entails that these attempts cannot lead to theories consistent with the invariance principle. Indeed, according to Prop. 4.1
if $\SREV$ is unitary and $\TREV$ anti-unitary, then the tern must belong either to ${\mathcal I}({\bf u})$ or to ${\mathcal I}({\bf d})$;
then the constraint stated by Prop. 5.1 forces the helicity to be zero.
%_____________________________________________________________________________________________ 6
\section{Theories based on ${\mathcal I}({\bf s})$}
According to Remark 5.1,
to search for possible theories of massless elementary particle with  non-zero helicity we have to turn on 
terns of ${\mathcal I}({\bf s})$.
In this section the features of these terns are investigated, and as a result we find also 
that this search is succesful.
In section 6.1 we prove that the component $U^-$ of a tern in ${\mathcal I}({\bf s})$ is constrained to be the
``mirrored version'' of $U^+$; in particular, $U^-$ is irreducible if and only if $U^+$ does.
The possible theories with $U^+$ irreducible and zero helicity are identified in section 6.2.1.
In section 6.2.2 it is proved that
for every non zero value  $m_+\in\ZZ$ there are two possible theories, each of them with opposite values
$\frac{m_+}{2}$ and $-\frac{m_+}{2}$ of the helicity.
%___________________________________________________________________________________________ 6.1
\subsection{Constraints between $U^+$ and $U^-$}
In a tern of ${\mathcal I}({\bf s})$, since $\sigma(\underline P)=S_0^+\cup S_0^-$, the projection operators 
$E^+$ and $E^-$ of Prop. 4.2 are both different from $\nop$, so that the representation $U$ of ${\mathcal P}_+^\uparrow$ 
is alsways reducible: $U=U^+\oplus U^-$, where $U^+=E^+UE^+$ and $U^-=E^-UE^-$ are the components of $U$ reduced
by ${\mathcal H}^+$ and ${\mathcal H}^-$ respectively. 
These reduced components, $U^+$ or $U^-$, can be reducible or not. The following proposition implies that
in a theory with transformer tern in ${\mathcal I}({\bf s})$ the reducibility of $U^+$ is equivalent to the reducibility of $U^-$.
\vskip.5pc\noindent
{\bf Proposition 6.1.}
{\sl
Let $(U,\SREV,\TREV)$ be a tranformer tern in
${\mathcal I}({\bf s})$, and let $F_+$ be a projection operator that reduces $U^+$; then the following statements hold.
\begin{itemize}
\item[(i)]
In the case that $\TREV$ is unitary, the projection operator $F_-^\trev=\TREV F_+\TREV$ reduces
$U^-$, and $F^\trev=F_++ F_-^\trev$ reduces $U$;
\item[(ii)]
in the case that $\SREV$ is anti-unitary, the projection operator $F_-^\srev=\SREV F_+\SREV$ reduces
$U^-$, and $F^\srev=F_++ F_-^\srev$ reduces $U$.
\item[(iii)]
$U^+$ is reducible if and only if $U^-$ is reducible.
\end{itemize}}
\vskip.4pc\noindent
{\bf Proof.}
We recall that if $T$ is a unitary or anti-unitary operator such that $TAT^{-1}=f(A)$, where $A$ is a self-adjoint operator with
spectral measure $E^A$ and $f$ is a continuous bijection of $\RR$, then $TE^A(\Delta)T^{-1}=E^A(f^{-1}(\Delta))$, for every 
Borel set $\Delta\subseteq\RR$.
\par
Now, if $\TREV$ is unitary, then
$\TREV^{-1}=\TREV$ and $\TREV P_0\TREV=-P_0$ follow from (11); this implies
$\TREV E^+\TREV=\TREV\chi_{[0,\infty)}(P_0)\TREV=\chi_{(-\infty,0]}(P_0)=E^-$.
If $F_+$ is a projection operator that reduces $U^+$, and hence $\nop<F_+< E^+$,
then
$F_-^\trev E^-=(\TREV F_+\TREV)E^-=(\TREV F_+\TREV)\TREV E^+\TREV=\TREV F_+ E^+\TREV=\TREV F_+ \TREV$ since $F_+< E^+$.
Therefore, $\nop< F_-^\trev< E_-$ is satisfied.
Now we show that
$[F_-^\trev,P_0^-]=[F_-^\trev,P_j^-]=[F_-^\trev,K_j^-]=[F_-^\trev,J_k^-]=\nop$, i.e. that
$F_-^\trev$ reduces $U^-$.
Since $P_0^-=E^-P_0E^-$ and $[F_+,P_0]=[F_+,P_0^+]=\nop$,
we have
$$\begin{array}{ll}
P_0^-F_-^\trev&=P_0^-F_-^\trev E^-=E^-P_0E^-\TREV F_+\TREV E^-=E^-P_0\TREV E^+\TREV \TREV F_+\TREV E^-\cr
&=-E^-\TREV P_0 E^+ F_+ \TREV E^-=
-E^-\TREV E^+ P_0 F_+ \TREV E^-\cr
&=-E^-\TREV E^+F_+ P_0\TREV E^-=-E^-\TREV  F_+ P_0\TREV E^-=
E^-\TREV F_+\TREV  P_0 E^-
\cr
&=
E^-F_-^\trev P_0 E^-=
F_-^\trev E^- P_0E^-=
F_-^\trev P_0^-.
\end{array}
$$
A similar derivation shows that
$[F_-^\trev,P_j]=[F_-^\trev,K_j^-]=[F_-^\trev,J_k^-]=\nop$; therefore $F_-^\trev$ reduces $U^-$.
Now we see that $F^\trev=F_++F_-^\trev$ reduces $U\mid_{{\mathcal P}_+}$. The equalities
$F^\trev P_0=(F_++F_-^\trev)P_0=P_0(F_++F_-^\trev)=P_0F^\trev$ immediately follow from
$P_0=E^+P_0 E^++E^-P_0E^-$ and $F_-^\trev E^-=F_-^\trev$, $F_+E^+=F_+$, $F_+E^-=F_-^\trev E^+=\nop$;
similarly,
$[F_-^\trev,P_j]=[F_-^\trev,J_k]=[F_-^\trev,K_j]=\nop$ hold. Hence, $F^\trev$ reduces $U\mid_{{\mathcal P}_+^\uparrow}$.
Moreover, 
$F^\trev\TREV=F_+\TREV+F_-^\trev\TREV=\TREV\TREV F_+\TREV+\TREV F_+\TREV\TREV=\TREV F_-^\trev+\TREV F_+=\TREV F^\trev$.
Therefore, $F^\trev$ reduces also $U\mid_{{\mathcal P}_+}$.
A quite similar argument proves statement (ii).
Statement (iii) is a direct consequence of statements (ii) and (ii).
\hfill{$\bullet$}
\vskip.5pc
So, in a tern of ${\mathcal I}({\bf s})$ the component $U^-$ is the ``mirrored'' version of $U^+$. 
According to Prop. 6.1,  ${\mathcal I}({\bf s})$ can be decomposed as 
${\mathcal I}({\bf s})={\mathcal I}_{irred}({\bf s})\cup {\mathcal I}_{red}({\bf s})$,
with obvious meaning of the notation.

%___________________________________________________________________________________________________ 6.2
\subsection{$U^+$ and $U^-$ irreducible}
In this section we derive theories based on terns in ${\mathcal I}_{irred}({\bf s})$. In these terns the irreducible 
component $U^+$ is identified, according to (16)-(18) in section 5.1, by a value $m^+$ of the parameter $m$;
analogously, $U^-$ is identified by the value $m^-$ of $m$, according to (19), (20).
Therefore, the Hilbert space of the representation $U=U^+\oplus U^-$ is ${\mathcal H}={\mathcal H}^+\oplus{\mathcal H}^-$,
 where
${\mathcal H}^+=L_2(\RR^3,d\nu)={\mathcal H}^-$.
If every vector $\psi=\psi^++\psi^-\in{\mathcal H}$, with $\psi^\pm\in{\mathcal H}^\pm$ is represented by 
the column vector $\psi\equiv\left[\begin{matrix}\psi^+\cr\psi^-\end{matrix}\right]$, then the generators are 
represented by the following matrices.
\vskip.4pc
$P_0=\left[\begin{matrix}p_0&0\cr 0&-p_0\end{matrix}\right]$,\quad $P_j=
\left[\begin{matrix}p_j&0\cr 0&p_j\end{matrix}\right]$,
,\hfill{(25.i)}
\vskip.4pc
$J_j=\left[\begin{matrix}J_j^{(0)}+\textsf j_j^+&0\cr 0&J_j^{(0)}+\textsf j_j^-\end{matrix}\right]$,\quad
$K_j=\left[\begin{matrix}K_j^{(0)}+\textsf k_j^+&0\cr 0&-(K_j^{(0)}+\textsf k_j^-)\end{matrix}\right]$,\hfill{(25.ii)}
\par\noindent
where $\textsf j_j^\pm$ and $\textsf k_j^\pm$ are given by (18) with the value $m^\pm$ of $m$.
Whenever $\psi^\pm\in{\mathcal H}^\pm$, the relation $\hat\lambda=\frac{m^\pm}{2}$ holds, of course.
\vskip.5pc\noindent
{\bf Proposition 6.2.}
{\sl If $U$ is a representation of a tern in ${\mathcal I}({\bf s})$ with generators given by (25) , then
$m^-=-m^+$; furthermore, whenever $m^+\neq 0$,  $\psi\in{\mathcal H}^\pm$ implies $\SREV\psi\in{\mathcal H}^\mp$.}
\vskip.4pc\noindent
{\bf Proof.}
Given any non vanishing vector $\psi\in{\mathcal H}^+$, let us define $\tilde\psi=\SREV\psi$.
Of course, $\tilde\psi=\varphi^++\varphi^-$, with $\varphi^\pm\in{\mathcal H}^\pm$.
By making use of (10) we find
$$
\hat\lambda\tilde\psi=\hat\lambda\SREV\psi=\SREV\SREV^{-1}\hat\lambda\SREV\psi=\SREV(-\hat\lambda)\psi=
-\frac{m^+}{2}\tilde\psi=-\frac{m^+}{2}\varphi^+
-\frac{m^+}{2}\varphi^-.
\eqno(26)
$$
On the other hand 
$\hat\lambda\tilde\psi=\hat\lambda(\varphi^++\varphi^-)=\frac{m^+}{2}\varphi^++\frac{m^-}{2}\varphi^-$,
hence by (26) we obtain
$$
-\frac{m^+}{2}\varphi^+-\frac{m^+}{2}\varphi^-=\frac{m^+}{2}\varphi^++\frac{m^-}{2}\varphi^-.\eqno(27)
$$
If $m^+\neq0$, then (22) implies $\varphi^+=0$, i.e. $\SREV\psi\in{\mathcal H}^-$, and $m^-=-m^+$.
\par
If $m^-\neq 0$, then the previous derivation, carried out starting from 
a vector $\psi\in{\mathcal H}^-$, leads to conclude that $\varphi^-=0$, i.e. $\SREV\psi\in{\mathcal H}^+$
and $m^+=-m^-$ hold.  
As a consequence,
$m^+= 0$ if and only if $m^-=0$.\hfill{$\bullet$}
\vskip.5pc
Thus, in  every tern of ${\mathcal I}({\bf s})$ with $U^+$ irreducible the operator $\SREV$ is identified by a matrix of the from
$$
\SREV=\left[\begin{matrix}0&\SREV_1\cr\SREV_2&0\end{matrix}\right],\eqno(28)
$$
and the representation
$U$  is characterized by a value of the parameter $m$ such that
$$
J_j=\left[\begin{matrix}J_j^{(0)}+\textsf j_j&0\cr 0&J_j^{(0)}-\textsf j_j\end{matrix}\right],\quad
K_j=\left[\begin{matrix}K_j^{(0)}+\textsf k_j&0\cr 0&-K_j^{(0)}+\textsf k_j\end{matrix}\right],\eqno(25.iii)
$$
where $\textsf j_j$ and $\textsf k_j$ are given by (18). Now we address the problem of determining the complete terns, 
that is to say the operators $\SREV$ and $\TREV$. Since 
$\sigma(\underline P)=S_0^+\cup S_0^-$, according to Prop.~4.1 the possible combinations are
\vskip.4pc\noindent
(U.U) $\SREV$ unitary and $\TREV$ unitary,
\par\noindent
(A.U) $\SREV$ anti-unitary and $\TREV$ unitary,
\par\noindent
(A.A) $\SREV$ anti-unitary and $\TREV$ anti-unitary.
%________________________________________________________________________________________________ 6.2.1
\subsubsection{Theories with zero helicity}
First we determine $\SREV$ and $\TREV$ for the terns with $m=0$, whose generators are given by 
\vskip.4pc\noindent
$P_0=\left[\begin{matrix}p_0&0\cr 0&-p_0\end{matrix}\right]$,\; $P_j=
\left[\begin{matrix}p_j&0\cr 0&p_j\end{matrix}\right]$,\;
$J_j=\left[\begin{matrix}J_j^{(0)}&0\cr 0&J_j^{(0)}\end{matrix}\right]$,\;
$K_j=\left[\begin{matrix}K_j^{(0)}&0\cr 0&-K_j^{(0)}\end{matrix}\right]$.\hfill{(29)}
\vskip.5pc\noindent
{\bf Combination (U.U).}
Let $(U,\SREV,\TREV)$ be a tern with generators given by (29) and $\SREV$ unitary. Let us define the unitary operator
$\hat S=\left[\begin{matrix}S_{11}&S_{12}\cr S_{21}&S_{22}\end{matrix}\right]=\Upsilon\SREV$, so that 
$\SREV=\Upsilon\hat S$. By making use of $\SREV P_j=-P_j\SREV$ in (5) and of the completeness
of $\bf P$ in $L_2(\RR^3,d\nu)$, we find that the entries $S_{mn}$  of $\hat S$ must be complex functions of ${\bf p}$:
$S_{mn}=S_{mn}({\bf p})$.
By making use of $[\SREV, P_0]=\nop$ in (5) we find $S_{12}=S_{21}=\nop$. Finally, $\SREV K_j=-K_j\SREV$ in (5) implies
$S_{11}({\bf p})=constant$, $S_{22}({\bf p})=constant$. The further condition 
$\SREV^2=\Id$ in (5) implies that there are two possibilities for $\SREV$:
$\SREV=\Upsilon\left[\begin{matrix}1&0\cr 0&1\end{matrix}\right]$ or $\SREV=\Upsilon\left[\begin{matrix}1&0\cr0&-1\end{matrix}\right]$.
\par\noindent
In order to determine $\TREV=\left[\begin{matrix}T_{11}&T_{12}\cr T_{21}&T_{22}\end{matrix}\right]$, we make use of the relations in (7), 
analogously to what done for determining $\SREV$ making use of the relations in (5). As a result we find
$\TREV=\left[\begin{matrix}0&1\cr 1&0\end{matrix}\right]$. 
Therefore, there are two inequivalent terns wth $\SREV$ unitary and $\TREV$ unitary.
\vskip.5pc\noindent
{\bf Combination (A.U).}
To determine the anti-unitary $\SREV$ we define the unitary operator
 $\hat S={\mathcal K}\SREV=\left[\begin{matrix}S_{11}&S_{12}\cr S_{21}&S_{22}\end{matrix}\right]$.
By making use of the relations (6) we find 
$\SREV={\mathcal K}\left[\begin{matrix}0&1\cr1&0\end{matrix}\right]$ when $\SREV^2=1$ and
$\SREV={\mathcal K}\left[\begin{matrix}0& 1\cr-1&0\end{matrix}\right]$ when $\SREV^2=-1$.
\par
The operator $\TREV$, being unitary, is $\TREV=\left[\begin{matrix}0&1\cr 1&0\end{matrix}\right]$ as in combination (U.U).
Therefore, also in this case there are two inequivalent terns.
\vskip.5pc\noindent
{\bf Combination (A.A).}
The operator $\SREV$, being anti-unitary, is $\SREV={\mathcal K}\left[\begin{matrix}0&1\cr1&0\end{matrix}\right]$ or
$\SREV={\mathcal K}\left[\begin{matrix}0& 1\cr-1&0\end{matrix}\right]$, as in combination (A.U).
To determine the anti-unitary $\TREV$ we define the unitary operator 
$\hat T={\mathcal K}\Upsilon\TREV=\left[\begin{matrix}T_{11}&T_{12}\cr T_{21}&T_{22}\end{matrix}\right]$.
By making use of the relations (8) we find 
$\TREV={\mathcal K}\Upsilon\left[\begin{matrix}1&0\cr0&e^{i\theta}\end{matrix}\right]$.
By transforming every operator $A$ into $WAW^{-1}$, where
$W=\left[\begin{matrix}1&0\cr0&e^{i\frac{\theta}{2}}\end{matrix}\right]$,
we obtain an equivalent theory. In so doing all generators are left invariant,
while $\TREV$ transforms into $\TREV={\mathcal K}\Upsilon\left[\begin{matrix}1&0\cr 0&1\end{matrix}\right]$.
So there are two inequivalent terns.
\vskip.5pc\noindent
Thus, in ${\mathcal I}({\bf s})$ there are six inequivalent terns with zero helicity and $U^+$ irreducible.
%_____________________________________________________________________________________________________ 6.2.2
\subsubsection{Terns with $m\neq 0$}
For the terns in ${\mathcal I}({\bf s})$ with representations chacterized by $m\neq 0$, the generators are
given by (25.i,iii). Equation (28) implies that $\SREV$ cannot be unitary; in this case, indeed, 
once introduced the unitary operator $\hat S=\Upsilon\SREV$, the relations 
$\SREV P_j=-P_j\SREV$ and $[\SREV,P_0]=\nop$ in (5) would imply
$\hat S=\left[\begin{matrix}S_1&0\cr0&S_2\end{matrix}\right]$ while (28) implies 
$\hat S=\left[\begin{matrix}0&S_1\cr S_2&0\end{matrix}\right]$, so that $\SREV$ should be $\nop$.
Therefore combinations (U.U) and (U.A) cannot occur. 
\vskip.5pc
Now we see that if $m\neq 0$ then $\TREV$ cannot be unitary.
Let $\TREV=\left[\begin{matrix}T_{11}&T_{12}\cr T_{21}&T_{22}\end{matrix}\right]$. Conditions $[P_j,\TREV]=\nop$ and
$P_0\TREV=-\TREV P_0$ in (7) imply $\TREV=\left[\begin{matrix}0&T_1({\bf p})\cr T_2({\bf p})&0\end{matrix}\right]$.
Now, the condition $[J_j,\TREV]=\nop$ in (7) implies $[J_j,T_1]=-2\textsf j_jT_1$ and 
$[J_j,T_2]=2\textsf j_jT_2$. In particular, for $j=3$ we have 
$$
i\left(p_2\frac{\partial T_n}{\partial p_1}-p_1\frac{\partial T_n}{\partial p_2}\right)=0,\; n=1,2.\eqno(30)
$$
On the other hand, the condition $K_j\TREV=-\TREV K_j$ in (7) leads to the equations 
$ip_0\frac{\partial T_1}{\partial p_j}=-2\textsf k_jT_1$ and $
ip_0\frac{\partial T_2}{\partial p_j}=2\textsf k_jT_2$; in particular
$$
\frac{\partial T_n}{\partial p_1}=-2i\frac{p_2p_3}{p_1^2+p_2^2}\frac{1}{p_0}T_n,\quad 
\frac{\partial T_n}{\partial p_2}=2i\frac{p_1p_3}{p_1^2+p_2^2}\frac{1}{p_0}T_n
.\eqno(31)
$$
By making use of (31) in (30) we obtain $2\frac{p_3}{p_0}T_n=0$, which can hold only if $T_n=0$.
This means that $\TREV$ cannot be unitary.
\vskip.5pc
Now we show that the combination (A.A), where both $\SREV$ and $\TREV$ are anti-unitary, can occur.
If $\SREV$ is anti-unitary, once defined 
$\hat S={\mathcal K}\SREV=\left[\begin{matrix}0&S_1\cr S_2&0\end{matrix}\right]$,
the relation $[\SREV,P_j]=\nop$ in (6) implies $S_1=S_1({\bf p})$ and $S_2=S_2({\bf p})$. The condition 
$[\SREV, K_j]=\nop$ of (6), where (16) is used, implies the equalities $S_1({\bf p})=constant$ and 
$S_2({\bf p})=constant$, which are consistent with the condition $[\SREV,P_0]=\nop$ in (6).
Hence we can set $\SREV={\mathcal K}\left[\begin{matrix}0&1\cr e^{i\theta}&0\end{matrix}\right]$.
The condition $\SREV^2=\pm1$ implies that $\SREV={\mathcal K}\left[\begin{matrix}0&1\cr 1&0\end{matrix}\right]$
if $\SREV^2=1$ and $\SREV={\mathcal K}\left[\begin{matrix}0&1\cr -1&0\end{matrix}\right]$
if $\SREV^2=-1$.
\par
Let $\TREV$ be anti-unitary. If we define $\hat T={\mathcal K}\Upsilon\TREV$, then conditions
$\TREV P_j=-P_j\TREV$ and $[P_0,\TREV]=\nop$ in (8) imply 
$\TREV={\mathcal K}\Upsilon  \left[\begin{matrix}T_1({\bf p})&0\cr 0&T_2({\bf p})\end{matrix}\right]$.
\par
By making use of this result and of (25.iii) in condition $[\TREV, K_j]=\nop$ in (8), we obtain
$T_1({\bf p})=constant$, $T_2({\bf p})=constant$, i.e.
$\TREV={\mathcal K}\Upsilon\left[\begin{matrix}1&0\cr 0&e^{i\theta}\end{matrix}\right]$.
By transforming every operator $A$ into $WAW^{-1}$, where
 $W=\left[\begin{matrix}1&0\cr 0&e^{-i\frac{\theta}{2}}\end{matrix}\right]$, an equvalent theory is obtained.
In so doing, all generators are left invariant whereas $\TREV$ is transformed into 
${\mathcal K}\Upsilon\left[\begin{matrix}1&0\cr 0&1\end{matrix}\right]$.
\vskip.5pc
Thus, we can conclude that for every $m\neq 0$ there are two inequivalent terns in ${\mathcal I}({\bf s})$, whose generators 
are given by (25.i) and (25.iii): a tern with 
$\SREV={\mathcal K}\left[\begin{matrix}0&1\cr 1&0\end{matrix}\right]$ and $\TREV={\mathcal K}\Upsilon\left[\begin{matrix}1&0\cr 0&1\end{matrix}\right]$,
and another tern with 
$\SREV={\mathcal K}\left[\begin{matrix}0&1\cr -1&0\end{matrix}\right]$ and
$\TREV={\mathcal K}\Upsilon\left[\begin{matrix}1&0\cr 0&1\end{matrix}\right]$.

%____________________________________________________________________________________ 6.3
\subsection{$U^+$ reducible}
The terns so far identified do not exhaust all possible irreducible terns. Indeed, 
the class of irreducible terns of ${\mathcal I}({\bf s})$ with $U^+$, and hence $U^-$ reducible is not empty.
Now we present an instance of these terns.
The Hilbert space of the tern is
${\mathcal H}={\mathcal H}^{(1)}\oplus {\mathcal H}^{(2)}\oplus {\mathcal H}^{(3)}\oplus {\mathcal H}^{(4)}$,
where
 ${\mathcal H}^{(n)}=L_2(\RR^3,d\nu)$, $n=1,2,3,4$.
Every vector $\psi\in{\mathcal H}$ is represented as a column vector
$\psi=\left[\begin{array}{c}\psi^{(3)}_+\cr\psi^{(3)}_-\cr \psi^{(5)}_+\cr\psi^{(5)}_-\end{array}\right]$, with $\psi^{(m)}_\pm\in L_2(\RR^3,d\nu)$.
\par\noindent
The projections $E^+$, $E^-$ and the self-adjoint generators satisfying (3) are
$$
E^+=\left[\begin{array}{cccc}1&0&0&0\cr 0&0&0&0\cr 0&0&1&0\cr 0&0&0&0\end{array}\right],\quad
E^-=\left[\begin{array}{cccc}0&0&0&0\cr 0&1&0&0\cr 0&0&0&0\cr 0&0&0&1\end{array}\right],\quad
P_0=\left[\begin{array}{cccc}p_0&0&0&0\cr 0&-p_0&0&0\cr 0&0&p_0&0\cr 0&0&0&-p_0\end{array}\right],
$$
$$
P_j=\left[\begin{array}{cccc}p_j&0&0&0\cr 0&p_j&0&0\cr 0&0&p_j&0\cr 0&0&0&p_j\end{array}\right],\quad
J_k=\left[\begin{array}{cccc}{\textsf j}_k&0&0&0\cr 0&{\textsf j}_k&0&0\cr 0&0&{\textsf j}_k&0\cr 0&0&0&{\textsf j}_k\end{array}\right],\quad
K_j=\left[\begin{array}{cccc}{\textsf k}_j&0&0&0\cr 0&-{\textsf k}_j&0&0\cr 0&0&{\textsf k}_j&0\cr 0&0&0&-{\textsf k}_j\end{array}\right].
$$
So, we have a {\sl reducible} representation of ${\mathcal P}_+^\uparrow$. Now we extend it to an {\sl irreducible} transformer tern,
by introducing a
unitary $\TREV$ and an anti-unitary $\SREV$ satisfying (5)-(9), as 
$\TREV=\left[\begin{array}{cccc}0&1&0&0\cr 1&0&0&0\cr 0&0&0&1\cr 0&0&1&0\end{array}\right]$ and
$\SREV={\mathcal K}\left[\begin{array}{cccc}0&0&0&1\cr 0&0&1&0\cr 0&-1&0&0\cr -1&0&0&0\end{array}\right]$.
\par\noindent
Indeed, let 
$A=\left[\begin{array}{cccc}A_{11}&A_{12}&A_{13}&A_{14}\cr A_{21}&A_{22}&A_{23}&A_{24}\cr 
A_{31}&A_{32}&A_{33}&A_{34}\cr A_{41}&A_{42}&A_{43}&A_{44}
\end{array}\right]$ be any self-adjoint operator of $\mathcal H$;
the conditions $[A,P_0]=[A,P_j]=[A,J_k]=[A,K_j]=[A,\TREV]=[A,\SREV]=\nop$ are satisfied if and only if
$A=\left[\begin{array}{cccc}a&0&0&0\cr 0&a&0&0\cr 0&0&a&0\cr 0&0&0&a
\end{array}\right]\equiv a\Id$ with $a\in\RR$. Thus the tern $(U,\SREV,\TREV)$ is irreducible.
%_________________________________________________________________________________________ 6.4
\subsection{Summary}
Let us summarize the class of  possible theories of massless particle based on irreducible terns with $r_0=0$.
\vskip.5pc\noindent
In ${\mathcal I}({\bf u})$ (resp. ${\mathcal I}({\bf d})$),\par\noindent
if $U$ is irreducible then 
there is just one tern, identified by (21) (resp, (22)), and $\hat\lambda=0$;\par\noindent
if $U$ is reducible, then each irreducible component is identical to the representation with zero helicity.
\vskip.5pc\noindent
In  ${\mathcal I}({\bf s})$, if $U^+$ is irreducible, then
\par\noindent
there are six possible theories with zero helicity, determined by (29), two for each
combination (U.U), (A,U), (A.A), according to section 6.2.1.
\vskip.4pc\noindent
For every $m\in\ZZ\setminus\{0\}$ there are  two terns with non-zero helicity, determined by (25.i,iii); they have both $\SREV$ ans $\TREV$ anti-unitary and opposite value of the hecity, $\pm\frac{m}{2}$, according to section 6.2.2.
In view of these results, whenever a non-zero value of the helicity is measured on an isolated system,
then its theory must be developed with one of these two new terns, while terns
with $\SREV$ unitary and $\TREV$ anti-unitary are inconsistent with the relativistic invariance principle.
As a conseqìence, also the opposite value must occur. For instance, the theory of the photon, which
experimentally exhibited $\pm 1$ helicity \cite{RamanAl}, must be based on one of the two terns with $m=\pm 2$.
%%________________________________________________________________________     7
\section{Identification of particle theories}
The identification of massless elementary particle theories can be addressed
by checking which of the tranformer terns identified in section 5,6  admit a position operator
 ${\bf Q}=(Q_1,Q_2,Q_3)$ according to section 3.3.

%_______________________________________________________________________________________________________  7.1
\subsection{Localizability of zero helicity systems}
Massless isolated systems with zero helicity whose theory is based on a tern of 
${\mathcal I}({\bf u})$ or ${\mathcal I}({\bf d})$
are always localizable \cite{NewtonWigner}, \cite{Jordan1}.
Following \cite{Jordan1}, indeed, if the representation $U$ of the tern in ${\mathcal I}({\bf u})$ is irreducible,
then  the Newton and Wigner three-operator
${\bf F}=(F_1,F_2,F_3)$, defined by 
$F_j=i\frac{\partial}{\partial p_j}-\frac{i}{2p_0^2}p_j$, is proved to be the unique three-operator such that (12) hold.
This conclusion holds also for the tern in ${\mathcal I}({\bf d})$.
Now we extend this result to a theory with transformer tern in ${\mathcal I}({\bf s})$.
%______________________________________________________________________________________  7.1.1
\subsubsection{Particle theories theories with terns in ${\mathcal I}({\bf s})$ and zero helicity}
According to section 6.1.1, there are six inequivalent terns in ${\mathcal I}({\bf s})$ with zero helicity and $U^+$ irreducible.
The Hibert space of these terns is
${\mathcal H}=L_2(\RR^3,d\nu)\oplus L_2(\RR^3,d\nu)$ and the generators are given by (29).
We check whether a position operator $\hat{\bf Q}$ with components
$\hat Q_j= \left[\begin{matrix}Q_{j11}&Q_{j12}\cr Q_{j21}&Q_{j22}\end{matrix}\right]$ exists such that the
conditions (12) hold. Hence, we introduce the operator
$\hat F_j=\left[\begin{matrix}F_j&0\cr 0&F_j\end{matrix}\right]$, 
where the $F_j$'s form the Newton and Wigner three-operator,
and the operators $\hat D_j=\hat Q_j-\hat F_j$.
\par
By making use of (29) it can be verified that $[F_j,P_k]=i\delta_{jk}$ and
$[J_j^{(0)},F_k]=i\hat\epsilon_{jkl}F_l$ hold. On the other hand the conditions (12.ii) hold, so that
$[\hat D_j,P_k]=i\delta_{jk}$ and
$[J_j^{(0)},\hat D_k]=i\hat\epsilon_{jkl}\hat D_l$.
These relations imply that $\hat D_j=\hat d(p_0)p_j$, where 
$\hat d(p_0)=\left[\begin{matrix}d_{11}(p_0)&d_{12}(p_0)\cr d_{21}(p_0)&d_{22}(p_0)\end{matrix}\right]$.
\par
Since $\SREV \hat F=-\hat F\SREV$ and $\TREV\hat F=\hat F\TREV$ hold for all six terns,  (12.i) implies that
the following relations must always hold.
$$
\SREV\hat D=-\hat D\SREV\,,\quad [\TREV,\hat D]=\nop\eqno(32)
$$
By making use of (32) in the six inequivalent terns we find that
\vskip.5pc\noindent
if $\SREV=\Upsilon\left[\begin{matrix}1&0\cr 0&1\end{matrix}\right]$ and 
$\TREV=\left[\begin{matrix}0&1\cr 1&0\end{matrix}\right]$ then
${\hat D}_j=\left[\begin{matrix}d_1(p_0)&d_2(p_0)\cr d_2(p_0)&d_1(p_0)\end{matrix}\right]p_j$;\hfill{(33.i)}
\par\noindent
if $\SREV=\Upsilon\left[\begin{matrix}1&0\cr 0&-1\end{matrix}\right]$ and 
$\TREV=\left[\begin{matrix}0&1\cr 1&0\end{matrix}\right]$ then
${\hat D}_j=\left[\begin{matrix}0&0\cr 0&0\end{matrix}\right]p_j$;\hfill{(33.ii)}
\par\noindent
if $\SREV={\mathcal K}\left[\begin{matrix}0&1\cr 1&0\end{matrix}\right]$ and 
$\TREV=\left[\begin{matrix}0&1\cr 1&0\end{matrix}\right]$ then
${\hat D}_j=\left[\begin{matrix}0&0\cr 0&0\end{matrix}\right]p_j$;\hfill{(33.iii)}
\par\noindent
If $\SREV={\mathcal K}\left[\begin{matrix}0&1\cr -1&0\end{matrix}\right]$ and 
$\TREV=\left[\begin{matrix}0&1\cr 1&0\end{matrix}\right]$ then
${\hat D}_j=\left[\begin{matrix}0&d(p_0)\cr d(p_0)&0\end{matrix}\right]p_j$;\hfill{(33.iv)}
\par\noindent
if $\SREV={\mathcal K}\left[\begin{matrix}0&1\cr 1&0\end{matrix}\right]$ and 
$\TREV={\mathcal K}\Upsilon\left[\begin{matrix}1&0\cr 0&1\end{matrix}\right]$ then
${\hat D}_j=\left[\begin{matrix}0&id(p_0)\cr -id_(p_0)&0\end{matrix}\right]p_j$;\hfill{(33.v)}
\par\noindent
if $\SREV={\mathcal K}\left[\begin{matrix}0&1\cr -1&0
\end{matrix}\right]$ and $\TREV={\mathcal K}\Upsilon\left[\begin{matrix}1&0\cr 0&1\end{matrix}\right]$
then
${\hat D}_j=\left[\begin{matrix}0&0\cr 0&0\end{matrix}\right]p_j$.\hfill{(33.vi)}
\vskip.5pc\noindent
Thus, there are three inequivalent  theories, identified by (33.ii), (33.iii) and (33.vi),
where there is a unique position operator $\hat{\bf Q}=\hat{\bf F}$ because $\hat{\bf D}=\nop$.
In particular, the theory corresponding to (33.iii) has $\SREV$ anti-unitary and $\TREV$ unitary.
%______________________________________________________________________________________ 7.2
\subsection{Non-localizability of non-zero helicity particles}
The several investigations about the localizability of massless particle, carried out through different approaches,
agree in concluding that massless particles with non-zero helicity are not localizable.
However, the theoretical structures for which the non-existence of a postion operator is proved
are terns where $\SREV$ is unitary and $\TREV$ is anti-iunitary \cite{Jordan1}.
In view of the analysis performed in the present work,
this is a serious shortcoming, because according to Prop. 4.1 these structure must be terns in ${\mathcal I}({\bf u})$
or ${\mathcal I}({\bf d})$. But in section 6 it is proved that irreducible terns with non-zero helicity
can exists only in ${\mathcal I}({\bf s})$. Therefore, these non-localizability proofs do not apply.
Now we present an alternative proof independent
of $\SREV$ and $\TREV$, so that it avoids the shortcomings of the previous ones.
\vskip.5pc
\noindent
{\bf Lemma 7.1.}
{\sl If $m\neq 0$ then the following system of equations has no solution for 
the functions $d_j({\bf p})$, $j=1,2,3$.
\vskip.5pc
$p_3\frac{\partial d_3}{\partial p_1}-p_1\frac{\partial d_3}{\partial p_3}=
-d_1+\frac{m}{2}\frac{p_2 p_3}{p_0(p_1^2+p_2^2)}$;\hfill{(d.1)}\vskip.5pc
$p_1\frac{\partial d_3}{\partial p_2}=p_2\frac{\partial d_3}{\partial p_1}$;\hfill{(d.2)}\vskip.5pc
$p_2\frac{\partial d_2}{\partial p_3}-p_3\frac{\partial d_2}{\partial p_2}=
-d_3-\frac{m}{2}\left(\frac{p_0p_1 p_2}{(p_1^2+p_2^2)^2}-\frac{p_1 p_2}{p_0(p_1^2+p_2^2)}\right)$;\hfill{(d.3)}\vskip.5pc
$p_3\frac{\partial d_1}{\partial p_1}-p_1\frac{\partial d_1}{\partial p_3}=
d_3-\frac{m}{2}\left(\frac{p_0p_1 p_2}{(p_1^2+p_2^2)^2}-\frac{p_1 p_2}{p_0(p_1^2+p_2^2)}\right)$;\hfill{(d.4)}\vskip.5pc
$p_2\frac{\partial d_1}{\partial p_3}-p_3\frac{\partial d_1}{\partial p_2}=
-\frac{m}{2}\left(\frac{p_0p_1^2}{(p_1^2+p_2^2)^2}-\frac{p_0}{p_1^2+p_2^2}-\frac{p_1 ^2}{p_0(p_1^2+p_2^2)}\right)$;\hfill{(d.5)}\vskip.5pc
$p_1\frac{\partial d_2}{\partial p_2}-p_2\frac{\partial d_2}{\partial p_1}=d_1$;\hfill{(d.6)}\vskip.5pc
$p_3\frac{\partial d_2}{\partial p_1}-p_1\frac{\partial d_2}{\partial p_3}=
-\frac{m}{2}\left(\frac{p_0p_2^2}{(p_1^2+p_2^2)^2}-\frac{p_0}{p_1^2+p_2^2}-\frac{p_2 ^2}{p_0(p_1^2+p_2^2)}\right)$;\hfill{(d.7)}\vskip.5pc
$p_1\frac{\partial d_1}{\partial p_2}-p_1\frac{\partial d_1}{\partial p_1}=d_2$;\hfill{(d.8)}\vskip.5pc
$p_2\frac{\partial d_3}{\partial p_3}-p_3\frac{\partial d_3}{\partial p_2}=
d_2+\frac{m}{2}\frac{p_3 p_1}{p_0(p_1^2+p_2^2)}$.\hfill{(d.9)}\vskip.5pc
}
\vskip.4pc\noindent
{\bf Proof.}
Once defined $\zeta({\bf p})={\bf p}\cdot{\bf d}({\bf p})=\sum_j p_jd_j({\bf p})$,
making use of (d.1)-(d.9)  we obtain, after a certain amount of computation, 
\vskip.4pc\noindent
$p_2\frac{\partial\zeta}{\partial p_3}-p_3\frac{\partial\zeta}{\partial p_2}=
\frac{m}{2}\frac{p_1p_0}{p_1^2+p_2^2}$\hfill{(34)}\par\noindent
$p_3\frac{\partial\zeta}{\partial p_1}-p_1\frac{\partial\zeta}{\partial p_3}=
\frac{m}{2}\frac{p_2p_0}{p_1^2+p_2^2}$\hfill{(35)}\par\noindent
$p_1\frac{\partial\zeta}{\partial p_2}-p_2\frac{\partial\zeta}{\partial p_1}=0$.\hfill{(36)
}\vskip.4pc\noindent
Now, equations (36) and (35) imply $\frac{\partial\zeta}{\partial p_3}=
-\frac{m}{2}\frac{p_0}{p_1^2+p_2^2}\frac{p_2}{p_1}+\frac{p_3}{p_2}\frac{\partial\zeta}{\partial p_2}$, 
which becomes\par\noindent
$\frac{m}{2}\frac{p_0}{p_1}=0$ by (34).\par\noindent
If $m$ were not zero, then the last equation would imply $p_0=0$. Therefore, there is no solution for $\zeta$, and hence for 
$d_j$.\hfill{$\bullet$}
\vskip.5pc\noindent
{\bf Propostition 7.1.}
{\sl If the a theory of a massless isolated system is based on a tern $(U,\SREV,\TREV)$ of non-zero helicity with $r_0=0$, 
then there is no position operator.}
\vskip.4pc\noindent
{\bf Proof.} 
If $U$ is irreducible then its tern must belong to ${\mathcal I}({\bf u})$ or ${\mathcal I}({\bf d})$; hence, according to Prop. 5.1
the helicity must be zero. Therefore, it is sufficient to prove the proposition for the case that $U$ is reducible.
If the helicity is not zero, then  an irreducible component $U^{(1)}$ must exists with non zero helicity; suppose that such $U^{(1)}$
belongs to ${\mathcal I}({\bf u})$, so that its Hilbert space is ${\mathcal H}^{(1)}=L_2(\RR^3,d\nu)$ and the generators are
given by (16)-(18). Hence
$U=U^{(1)}\oplus U^{(2)}$, where $U^{(2)}$ can be reducible.
The Hilbert space of $U$ decomposes as  ${\mathcal H}= {\mathcal H}^{(1)}\oplus{\mathcal H}^{(2)}$, where ${\mathcal H}^{(2)}$
is the Hilbert space of $U^{(2)}$.
According to such a decomposition, the generators are
\vskip.4pc
$P_0=\left[\begin{matrix}p_0&0\cr 0&P^{(2)}_0\end{matrix}\right]$,\quad $P_j=
\left[\begin{matrix}p_j&0\cr 0&p_j\end{matrix}\right]$,
\hfill{(37.i)}
\vskip.4pc
$J_j=\left[\begin{matrix}J_j^{(0)}+\textsf j_j&0\cr 0&J^{(2)}_j\end{matrix}\right]$,\quad
$K_j=\left[\begin{matrix}K_j^{(0)}+\textsf k_j&0\cr 0&K_j^{(2)}\end{matrix}\right]$,\hfill{(37.ii)}.
\vskip.5pc\noindent
If a position operator $\hat{\bf Q}=(\hat Q_1,\hat Q_2,\hat Q_3)$ exists, 
then each coordinate operator $\hat Q_j$ must have the form
$\hat Q_j= \left[\begin{matrix}Q_{j11}&Q_{j12}\cr Q_{j21}&Q_{j22}\end{matrix}\right]$,
where $Q_{j11}$ is a  self-adjoint operator of $L_2(\RR^3,d\nu)$.
Let us define the self-adjoint operators
$$
\hat{D}_j=\left[\begin{matrix}Q_{j11}-F_j&Q_{j12}\cr Q_{j21}&Q_{j22}-F_j\end{matrix}\right]
\equiv\left[\begin{matrix}d_{j}&D_{j12}(p_0)\cr D_{j21}&D_{j22}\end{matrix}\right],
$$ 
where $ F_j$ is the j-th component of the Newton and Wigner operator.
We prove that (12) implies conditions for $d_j$ that cannot be satisfied, and thus no position operator exists.
\par
Since $[ F_j,p_k]=i\delta_{jk}$ holds, (12.iii) implies $[\hat D_j,P_k]=\nop$, in particular
$[d_j,p_k]=\nop$;
therefore each $d_j$ is a function of $\bf p$: $d_j=d_j({\bf p})$.
Since $[J_j^{(0)}, F_k]=i\hat\epsilon_{jkl}F_l$, (12.iii) implies 
$[J_j^{(0)}+\textsf j_j, d_k({\bf p})]=i\hat\epsilon_{jkl} d_l({\bf p})-[\textsf j_j,F_k]$,
i.e., 
$$
[J_j^{(0)},d_k({\bf p})]=i\hat\epsilon_{jkl} d_l({\bf p})-[\textsf j_j,F_k].\eqno(38)
$$
Making use of  $J_j^{(0)}=-i\left(p_k\frac{\partial}{\partial p_l}-p_l\frac{\partial}{\partial p_k}\right)$ and (18), the computation of
$[\textsf j_j,F_k]$ for all $(j,k)$ yields equations (d.1)-(d.9). Lemma 7.1. proves that no solution ${\bf d}({\bf p})$ exists
for these equations if $m\neq 0$. Thus, no position operator can exist.\hfill{$\bullet$}
\vskip.5pc
It must be stressed how this proof of non-localizability is carried out without  making use of 
the operators $\SREV$ and $\TREV$, so that it holds also if space inversion or time reversal are not assumed
to be symmetries of the system.
%%%%%%%%%%%%%%%%%%%%%%%%%%%%%%%%%%%%%%%%%%%%%%%%%%%%%
%%%%%%%%%%%%%%%%%%%%%%%%%%%%%%%%%%%%%%%%%%%%%%%%%%%% 
                   %%%%           BIBLIOGRAPHY
%%%%%%%%%%%%%%%%%%%%%%%%%%%%%%%%%%%%%%%%%%%%%%%%%%%%%%

\end{document}